\documentclass[fleqn,usenatbib]{mnras}

\usepackage{newtxtext,newtxmath}

\usepackage[T1]{fontenc}
\usepackage{ae,aecompl}

\usepackage{url}

\usepackage{graphicx}	
\usepackage{amsmath}	
\usepackage{subcaption}
\usepackage{multirow}
\usepackage{afterpage}
\usepackage{lscape}
\usepackage{lineno}
\usepackage{color,soul}

\title[Optimising magnitude-limited training]{Optimising a magnitude-limited spectroscopic training sample for photometric classification of supernovae}

\author[J. Carrick et al.]{
Jonathan E. Carrick,$^{1}$\thanks{E-mail: j.carrick@lancaster.ac.uk}
Isobel M. Hook,$^{1}$ Elizabeth Swann$^2$, Kyle Boone$^{3}$, \newauthor Chris Frohmaier$^2$, Alex G. Kim$^4$, Mark Sullivan$^5$, (The LSST Dark \newauthor Energy Science Collaboration)
\\
$^{1}$Physics Department, Lancaster University, Bailrigg, Lancaster, LA1 4YB, UK\\
$^{2}$Institute of Cosmology and Gravitation, University of Portsmouth, PO1 3FX, UK\\
$^{3}$DIRAC Institute, Department of Astronomy, University of Washington, 3910 15th Ave NE, Seattle, WA, 98195, USA\\
$^{4}$Physics Division, Lawrence Berkeley National Laboratory, Berkeley, CA 94720, USA\\
$^{5}$School of Physics and Astronomy, University of Southampton, Highfield, Southampton, SO17 1BJ, UK
}

\date{Accepted XXX. Received YYY; in original form ZZZ}

\pubyear{2021}

\begin{document}
\label{firstpage}
\pagerange{\pageref{firstpage}--\pageref{lastpage}}
\maketitle

\begin{abstract}
In preparation for photometric classification of transients from the Legacy Survey of Space and Time (LSST) we run tests with different training data sets. Using estimates of the depth to which the 4-metre Multi-Object Spectroscopic Telescope (4MOST) Time Domain Extragalactic Survey (TiDES) can classify transients, we simulate a magnitude-limited sample reaching $r_{\textrm{AB}} \approx$ 22.5~mag. We run our simulations with the software \textsc{snmachine}, a photometric classification pipeline using machine learning. The machine-learning algorithms struggle to classify supernovae when the training sample is magnitude-limited, in contrast to representative training samples. Classification performance noticeably improves when we combine the magnitude-limited training sample with a simulated realistic sample of faint, high-redshift supernovae observed from larger spectroscopic facilities; the algorithms' range of average area under ROC curve (AUC) scores over 10 runs increases from 0.547--0.628 to 0.946--0.969 and purity of the classified sample reaches 95 per cent in all runs for 2 of the 4 algorithms. By creating new, artificial light curves using the augmentation software \textsc{avocado}, we achieve a purity in our classified sample of 95 per cent in all 10 runs performed for all machine-learning algorithms considered. We also reach a highest average AUC score of 0.986 with the artificial neural network algorithm. Having `true' faint supernovae to complement our magnitude-limited sample is a crucial requirement in optimisation of a 4MOST spectroscopic sample. However, our results are a proof of concept that augmentation is also necessary to achieve the best classification results.

\end{abstract}




\section{Introduction}
\label{sec:intro}
In this era of big data, new challenges are being presented in the astronomical community, some of which have the potential to be solved using machine learning. The Vera C. Rubin Observatory's Legacy Survey of Space and Time (LSST)\footnote{\url{https://www.lsst.org/}} is expected to discover 3--4 million supernovae over its 10-year survey. This unprecedented and vast accumulation of optical transient data means, however, that with current spectroscopic facilities and their capabilities it is unrealistic to follow up every discovery for classification. Type Ia supernovae play a key role in cosmology as they are standardisable. After applying corrections for light curve shape and colour, and also host-galaxy properties, they exhibit very similar peak luminosity. Type Ia supernovae therefore provide an excellent standardisable candle with which to measure the Universe's accelerated expansion \citep{1998AJ....116.1009R, 1999ApJ...517..565P}. With the number of Type Ia supernova discoveries from LSST, we will be able to test cosmological models and constrain parameters, such as the dark energy equation of state, to a much higher degree of precision than from any previous dataset. To use supernovae as cosmological probes, we first need to be sure that they are in fact Type Ia. Supernova type is traditionally determined by the chemical signatures that appear in their spectra, for example the presence of silicon in Type Ia supernovae \citep{1997ARA&A..35..309F}. Not wanting to waste the potential supernova science of all these objects, we therefore need to consider other methods of classification for the transient events that are not spectroscopically followed-up. Hence, photometric classification of supernovae using machine learning provides a solution.

Photometric classification with machine learning is a process that takes supernova light curve observations, generally with multiple filters, and determines the supernova type based on information learnt from a given training sample of supernova light curves with confirmed type. In preparation for LSST and other future surveys, there has recently been a great focus into what makes a good training sample for photometric classification of supernovae. As with many typical machine learning problems, a training sample that is representative of the whole dataset that is to be classified -- the `test set' or `target sample' -- seems a necessity \citep{2016ApJS..225...31L, 2017ApJ...837L..28C, 2019MNRAS.483....2I, 2020MNRAS.491.4277M, 2019PASP..131k8002M}. A representative training sample is one whose feature-space distributions are similar to those of the test set. Machine-learning models trained on samples which are representative of the target distribution are expected to perform well in classification tasks, so long as they have sufficient coverage of the test data \citep{2017MNRAS.468.4323B}. For supernovae, there are broad variations in their light curves across many magnitudes and redshifts. A representative training sample should include the features associated with these variations. 

None the less, work into data augmentation methods shows that focusing on accumulating a spectroscopic sample of supernovae that is fully representative may not be necessary. As long as one starts with a sample that has reasonable coverage of the full test set, augmentation can fill the gaps to create a much more representative training sample. Using Gaussian processes to model supernova light curves, it is possible to create new simulated light curves that cover more of the test set feature-space and add them into the training sample, making it artificially more representative. This approach is used in the works by \cite{2018MNRAS.473.3969R} and \cite{2019AJ....158..257B}, yielding very promising classification results. The latter of these was the winning solution to the Photometric Light Curve Astronomical Time-Series Classification Challenge\footnote{\url{https://www.kaggle.com/c/PLAsTiCC-2018}} (PLAsTiCC; results of the challenge are discussed in Hlo{\v{z}}ek et al., in preparation), which required classifying simulated LSST data using a provided non-representative training sample. The training sample mimicked a real set of light curves (of many types of object, not just supernovae) with spectroscopically-confirmed type and a preference to brighter, low-redshift objects. With augmentation to create artificial light curves and help cover the whole feature-space, less time is required from spectroscopic resources to build a faint training sample.

Given the constraints on observing resources for spectroscopic follow-up, we set out to determine how these limited resources would be best used, i.e. how to get the best resulting photometric classification of the remaining sample. In particular we consider the use of the 4MOST (4-metre Multi-Object Spectroscopic Telescope\footnote{\url{https://www.4most.eu/cms/}}) spectrograph, which will carry out the Time-Domain Extragalactic Survey (TiDES, \citealt{2019Msngr.175...58S}), a campaign for spectroscopic follow-up. The follow-up potential with 4MOST is determined by its survey overlap (both angular and temporal) with LSST's observing strategy, its cadence, and TiDES' allocated 250000~fibre-hours. 4MOST will be conducting multiple surveys with different science goals simultaneously \citep{2019Msngr.175....3D}. TiDES will be `piggy-backing' on other surveys and will not be driving where to point 4MOST or for how long. We assume 1~h field visits based on \cite{2020MNRAS.497.4626T}. For TiDES, a field visit exposure time of 1~h in combination with a spectral success criterion (SSC) effectively imposes a magnitude limit to spectroscopically confirmed supernovae. Using the 4MOST capabilities as a guide, we set out to optimise a spectroscopic training sample of supernovae. 

We also consider the role of redshift in the photometric classification of supernovae. For Type Ia cosmology we require spectroscopic redshifts of supernovae, as cosmology with photometric redshifts will be skewed and is prone to contamination \citep{2019PhRvD.100d3542L, 2021PhRvD.103b3524M}. At the end of the TiDES survey, we will have a spectroscopically confirmed sample of supernovae that will be used as the basis of our training sample. We will also have spectroscopic redshifts for many host galaxies of LSST supernovae for which we do not have a classification. These are the supernovae that we will want to photometrically classify for cosmology. Spectroscopic redshifts are necessary for cosmology, but can also be used as an additional feature in our classifiers. \cite{2016ApJS..225...31L} concluded that including photometric redshifts of supernova host galaxies does not have a significant impact on classification when using representative training samples, although the level of accuracy is model- and algorithm-dependent. We investigate the three cases of using spectroscopic, photometric and no redshift in classification.

Section \ref{sec:surveys} introduces the context of our work in future supernova surveys. Section \ref{sec:ML} explores the machine learning methods used in our photometric classification and we discuss representative training samples and the role of redshift. In Section \ref{sec:training} we present the simulations of a 4MOST-based training sample. Using this training sample we look at results of photometric classification in Section \ref{sec:results}, and explore and discuss methods of improving these results. Our findings are summarised in Section \ref{sec:conclusions}.

\section{Future ground-based supernova surveys}
\label{sec:surveys}
The Rubin Observatory will revolutionise astronomical sky surveys due to its large primary mirror (diameter of 8.4~m) and wide field of view (9.6~deg$^2$), and its immense data stream, gathering $\sim$20~TB of data per night and covering the visible night sky every 3--4 nights. It will carry out LSST, using filters $u$, $g$, $r$, $i$, $z$ and $y$, spanning the ultraviolet to near-infrared. Its Wide-Fast-Deep (WFD) survey will cover the majority of the southern sky (18000~deg$^2$), reaching up to redshift $z \sim$ 0.8 for supernova discovery (discounting superluminous supernovae), where specific depth will depend upon survey strategy. In addition to the WFD survey, LSST's Deep-Drilling-Fields (DDFs) include at least 4 patches of sky that will be visited more often and therefore reaching deeper coadded magnitudes. The details of observing strategy are still being reviewed (discussion can be found in \citealt{2017arXiv170804058L}).

Despite not being able to follow up every transient event from LSST, TiDES will obtain as many spectra as possible for the purposes of cosmology and creating a basis for our training sample. 4MOST, an instrument of the European Southern Observatory\footnote{\url{https://www.eso.org/public/}} (ESO), is particularly well-suited for this task, with first light expected in 2023. It will be installed on the Visible and Infrared Survey Telescope for Astronomy\footnote{\url{http://www.eso.org/sci/facilities/paranal/telescopes/vista.html}} (VISTA) in Chile, at a similar latitude to the Rubin Observatory.

Before we can assess the potential success of our science goals, we have to first consider the practical capabilities of 4MOST. 4MOST-TiDES is one of ten consortium surveys \citep{2019Msngr.175....3D}, each with its own individual objectives. In the context of this paper, we particularly consider TiDES' science goals (i) spectroscopic classification of live transients (TiDES-SN) and (ii) spectroscopy of supernova host galaxies (TiDES-Hosts) \citep{2019Msngr.175...58S}.

Once 4MOST's survey strategy is finalised, TiDES will need to decide how best to distribute its allocated 250000 fibre-hours of spectroscopy. TiDES will be exploiting the fact that wherever 4MOST points in the extragalactic sky, there will be LSST live transients that we want to follow up. Hence, rather than driving the 4MOST pointings, TiDES will be `piggy-backing' on the other surveys as the target density of transients is not high enough for efficient observations on its own; TiDES utilises approximately 2 per cent of 4MOST fibres (30--35 low-resolution spectrograph fibres), so it would not be efficient to use 4MOST exclusively for LSST transients. Once receiving LSST transient alerts/detections, TiDES will aim for a rapid turnaround time of 3--4 days in which to target the allocated fibres on to these objects and obtain their spectra.

We estimate that TiDES will be able to classify transient spectra to magnitudes as faint as $r_{\textrm{AB}} \approx 22.5$~mag. We explain the origin and implications of this magnitude limit in Section \ref{subsec:maglimit}. It will be the main factor influencing the training sample of supernovae we expect to produce using 4MOST. LSST is expected to detect transients fainter than this, making point-source detections down to a depth of $r_{\textrm{AB}} \approx 24$~mag in a single field visit. Consequently, the performance of our classification algorithms depends on how we deal with this magnitude limit.

TiDES will target all live transients ($r_{\textrm{AB}}$~<~22.5~mag) in each 4MOST pointing during grey and dark time. Depending on the nature of the final LSST cadence, we expect a density of 6--12 live transients per pointing. Over the 5-year duration of TiDES this equates to an expected total of >30000 transients, with the remaining fibre-hours used to measure host-galaxy redshifts of LSST transients. Final numbers are highly dependent on both LSST and 4MOST survey strategies that at the time of writing are not yet finalised. The survey strategy, transient populations and cosmological constraints expected from TiDES will be presented in future works (Frohmaier et al., in preparation). TiDES' spectroscopic sample can be used for training our machine-learning algorithms to subsequently classify other LSST transients. The supernova light curves that we will photometrically classify are those for which we have secured host-galaxy redshifts. Combining the Type Ia in the spectroscopic and photometrically classified samples, altogether, TiDES therefore expects to produce the largest cosmological sample of Type Ia supernovae by over an order of magnitude.

Classifying live supernovae that are fainter than 4MOST's limit would require use of 8-m and larger telescope facilities, such as the Very Large Telescope\footnote{\url{https://www.eso.org/public/teles-instr/paranal-observatory/vlt/}} (VLT) and the upcoming, next generation Extremely Large Telescope\footnote{\url{https://www.eso.org/public/teles-instr/elt/}} (ELT), Thirty Meter Telescope\footnote{\url{https://www.tmt.org/}} (TMT) and Giant Magellan Telescope\footnote{\url{https://www.gmto.org/}} (GMT). However, to classify live supernovae, time on these telescopes is likely to be even more limited than on 4MOST, so we do not expect more than a few hundred sources to be observed. We return to this in Section \ref{subsec:fainter}.

\section{Photometric classification of supernovae with machine learning}
\label{sec:ML}

For the task of photometric classification of supernovae with machine learning, we opted to use \textsc{snmachine} \citep{2016ApJS..225...31L}, a classification pipeline available through the Rubin Observatory LSST Dark Energy Science Collaboration\footnote{\url{https://lsstdesc.org/}} (DESC).

All our machine-learning simulations so far have been conducted using the Supernova Photometric Classification Challenge (SPCC) dataset (we use the simulations that were updated following the original challenge, \citealt{2010arXiv1001.5210K, 2010PASP..122.1415K}). The data are simulated light curves of 21319 supernovae of different types (Ia, Ib, Ic, Ibc, II, IIP, IIL and IIn)\footnote{The proportions of these types, grouped by Ia, Ibc and II, can be seen where we present class balance in appendix \ref{app:proportions}}. The light curves have been simulated to mimic Dark Energy Survey (DES) observations, using the filters $g$, $r$, $i$ and $z$. LSST has additional filters $u$ and $y$, which may improve classification, although is close to the SPCC as LSST's supernova cosmology focus will be on the $g$, $r$, $i$, $z$ bands (\citealt{2018arXiv180901669T} finds that filters $u$ and $y$ provide negligible cosmological information), with a similar cadence of observations every few days in each filter. The light curves consist of flux measurements and associated uncertainties in the four bands at times specified by the Modified Julian Date. In \textsc{snmachine}, the light curves are aligned such that they all start at time $t=0$.

In this work we primarily consider the binary classification of Type Ia vs. non-Ia (positive vs. negative), due to our focus on applications to Type Ia cosmology. However, we also run a few tests in which \textsc{snmachine} returns a classification probability for each supernova being either a Type Ia, Ibc (Ib, Ic, Ibc) or II (II, IIP, IIL, IIn), as in, e.g. \cite{2020MNRAS.491.4277M} and also the many solutions to the SPCC and PLAsTiCC challenges. In this case, we still apply a binary Ia vs. non-Ia classification, but with the aim of investigating whether considering Type Ibc and Type II light curves separately in the training would reduce the number of false positives (non-Ia light curves classified as Ia). We return to this in Section \ref{subsec:3class}.

The process for classification starts with extracting features from all the supernova light curves in the dataset. We use the wavelet decomposition method implemented in \textsc{snmachine} that extracts the wavelet coefficients that parametrize the light curves of each supernova using a Gaussian process regression. Using 100 points on the Gaussian process curve and a two-level wavelet transform, the output of wavelet decomposition consists of a highly redundant 1600 (400 per filter) coefficients per supernova. To reduce the dimensionality whilst preserving the useful information, a principal component analysis (PCA, \citealt{doi-10.1080-14786440109462720,HotellingPCA}) is applied. After PCA, there are 20 features per light curve. This was chosen as \cite{2016ApJS..225...31L} finds that reducing the number of features from 1600 to 20 using PCA retains 98 per cent of the dataset's information. If redshift is included as an additional feature, we add this to the feature set for each supernova, making a total of 21 features. We discuss the role of redshift in a representative training sample in Section \ref{subsec:representative} and the use of redshift in our 4MOST sample simulations in Section \ref{subsec:redshift}.

\textsc{snmachine}'s machine-learning classification algorithms are trained to associate feature values with the chosen classes (e.g. Ia and non-Ia) from supernovae in the training sample. When presented with the test set light curves, \textsc{snmachine} returns a probability of each supernova being either Type Ia or non-Ia. The classification algorithms are k-nearest neighbours (KNN), support vector machines (SVM), artificial neural networks (ANN) and boosted decision trees (BDT) explained in detail in \cite{2016ApJS..225...31L}. We present results from all four, although our primary focus in this paper is how different training samples affect classification results. For our implemented case of wavelet decomposition feature extraction, \textsc{snmachine}'s naive Bayes algorithm performs barely better than randomly in classification (even in the case of a representative training sample) and is therefore disregarded.

We determine which supernovae to use for our training sample independently to the \textsc{snmachine} classification pipeline. The remaining supernovae in the dataset are then used as the test set. Hence, we do multiple runs in our simulations to try to understand any trends in the classification results between different types of training sample.

As we use only the SPCC dataset in the machine-learning aspect of our tests (we use a separate mock catalogue of LSST supernovae to determine our 4MOST magnitude limit, discussed in Section \ref{subsec:maglimit}), which is considerably smaller than the sample we expect LSST and 4MOST to produce (particularly at bright magnitudes), our results should be considered as a proof of concept, rather than a definitive outcome for any dataset. 

\begin{figure*}
    \begin{subfigure}{.45\textwidth}
        \includegraphics[width=1\linewidth]{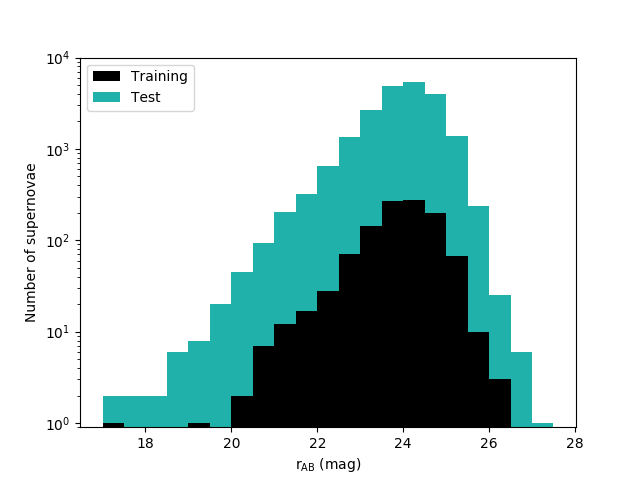}
        \caption{Stacked magnitude histogram of the training and test sets}
        \label{subfig:mag_hist_representative}
    \end{subfigure}
    \begin{subfigure}{.45\textwidth}
        \vspace{0.3cm}
        \includegraphics[width=1\linewidth]{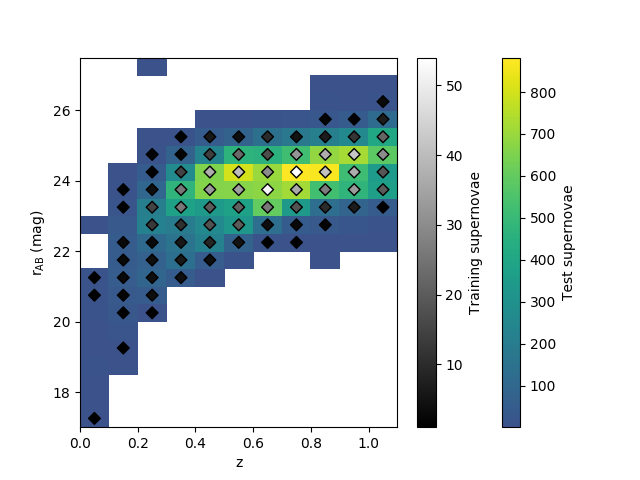}
        \caption{2D histogram of the relative distributions of redshift and magnitude in the training and test sets}
        \label{subfig:z-mag_representative}
    \end{subfigure}
    \caption{A random sample of 1103 supernovae chosen for training follows similar distributions of magnitudes and redshifts in the test set. Note that the single faint supernova ($r_{\textrm{AB}}>27.5$), which appears anomalous to the rest of the dataset, is a result of this particular simulated light curve only having two very faint observations in the $r$-band.}
    \label{fig:representative}
\end{figure*}

\begin{figure}
    \centering
    \includegraphics[width=1\linewidth]{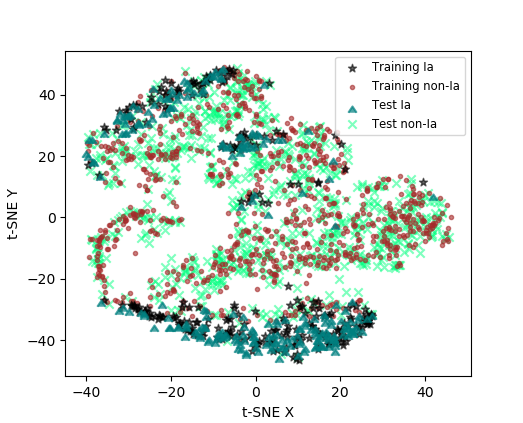}
    \caption{A t-SNE plot showing a 2D representation of the 21-dimensional feature-space after PCA and including spectroscopic redshift. Type Ia and non-Ia supernovae are found in their own respective clusters and regions of the plot. A randomly drawn training sample has supernovae of the same types occupying the same feature-space as those in the corresponding test set. This plot only includes one tenth of the test set for clarity.}
    \label{fig:tsne-representative}
\end{figure}

\subsection{Metrics}
\label{sec:metrics}

Receiver Operator Characteristic (ROC) curves compare the True Positive Rate (TPR, a.k.a. completeness) against False Positive Rate (FPR, a.k.a. contamination) for a range of probability thresholds. TPR and FPR are defined as

\begin{equation}
    \textrm{TPR} = \frac{\textrm{TP}}{\textrm{TP} + \textrm{FN}},
    \label{eq:tpr}
\end{equation}

\begin{equation}
    \textrm{FPR} = \frac{\textrm{FP}}{\textrm{FP} + \textrm{TN}},
    \label{eq:fpr}
\end{equation}

\noindent where TP is the number of true positives (Ia classified as Ia), FP is the number of false positives (non-Ia classified as Ia), TN is the number of true negatives (non-Ia classified as non-Ia) and FN is the number of false negatives (Ia classified as non-Ia).

For each run with \textsc{snmachine} we produced ROC curves for each machine-learning algorithm. A ROC curve's AUC value equals 1 for a perfect classifier (TPR~=~1 and FPR~=~0) and 0.5 for a completely random classifier.

In the context of using Type Ia supernovae for cosmology, it is crucial that our classified sample has very low contamination and so we also consider purity as an essential metric. Purity is defined as

\begin{equation}
    \textrm{Purity} = \frac{\textrm{TP}}{\textrm{TP} + \textrm{FP}}.
    \label{eq:purity}
\end{equation}

In general, there is a trade-off between completeness and purity.

For any classification problem, the measure of success depends on the choice of metric. For increasingly large datasets, e.g. from LSST, there will come a point at which systematic error dominates over statistical error\footnote{Statistical error increases by $\sqrt{N}$, where $N$ is the number of objects in the dataset, whereas contamination rate caused by systematic error is proportional to $N$.}. Therefore, we assume that we are above the completeness level at which contamination of our classified sample from systematic effects dominates statistical error and we set a high target purity value of 95 per cent. An in-depth look into when exactly this occurs for LSST requires further studies.

\subsection{A representative training sample}
\label{subsec:representative}
Before running \textsc{snmachine} with our simulated 4MOST training sample, we first follow the procedure from \cite{2016ApJS..225...31L} to demonstrate what is possible when using representative training samples. First, we discuss what we mean by `representative'.

In a given dataset with well-defined classes, a randomly drawn training sample of sufficient size has proportions of different supernova types equal to those in the test set (this is presented in appendix \ref{app:proportions} -- see Fig. \ref{subfig:fracs-representative} -- with a brief discussion on class balance). It is blind to supernova light curve parameters and has similar distributions in magnitude and redshift, shown in Fig. \ref{fig:representative}. Consequently, a randomly drawn training set samples the full range of feature values existing in the test set. To illustrate this, we show a two-dimensional representation of the 21 wavelet features (after PCA and including spectroscopic redshift), separated into training and test sets, and also by type (Ia vs. non-Ia). This was done by adopting t-distributed stochastic neighbour embedding (t-SNE, \citealp{vanDerMaaten2008}), a method that clusters similar high-dimensional objects together. Clear separation between classes is indicative of intrinsic differences in their respective features and suggests that accurate classifcation is possible. Shown in Fig. \ref{fig:tsne-representative} for a randomly drawn training sample, training and test Type Ia supernovae occupy the same regions of feature-space, and similarly for training and test non-Ia supernovae. Hence, given sufficient size, a randomly drawn training sample is representative of the corresponding test set, and, for the rest of this paper, we therefore refer to a randomly selected training sample as being representative, as in \cite{2016ApJS..225...31L}. In our tests, representative training samples are created by taking a random selection of 1103 objects from the SPCC. This is the same size as the original sample in the classification challenge.

\begin{figure*}
    \includegraphics[width=\textwidth]{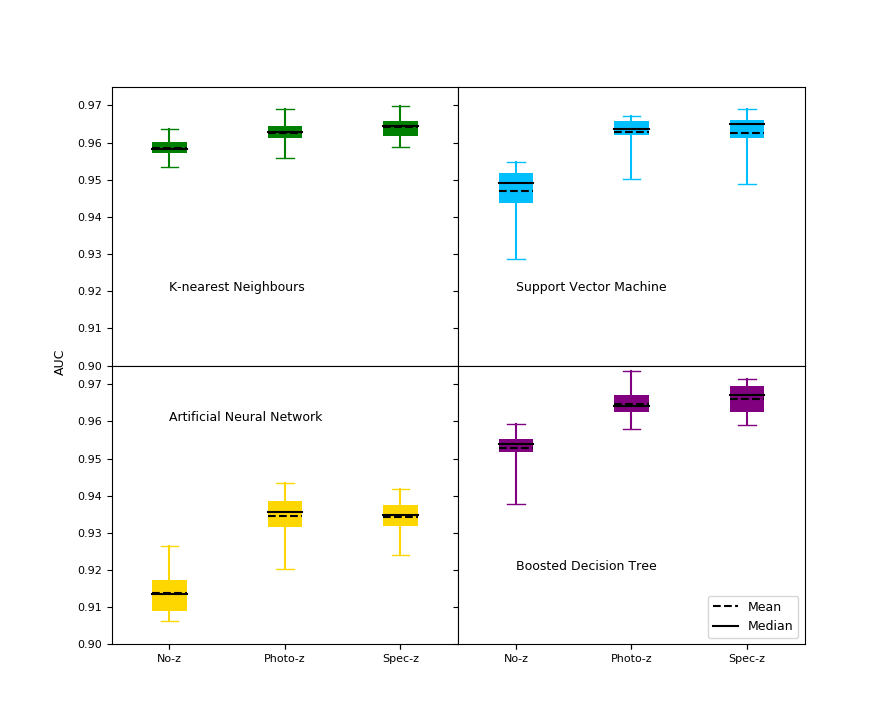}
    \caption{Boxplots showing the AUC scores over 20 classification runs for representative training samples comparing the use of no redshift information (No-z), photometric redshifts (Photo-z) and spectroscopic redshifts (Spec-z).}
    \label{fig:representative-aucs}
\end{figure*}

We compare using the same training sample in an individual run, but with either the `true' redshift\footnote{This is the \texttt{SIM\_REDSHIFT} parameter in the header of each supernova file.}, a photometric redshift or no redshift information used in both training and test samples to investigate which case is most successful for classification. The `true' redshift is used to mimic a spectroscopic redshift and is defined as such from this point onwards. The AUC scores over 20 runs are shown as boxplots in Fig. \ref{fig:representative-aucs} and summarised in Table \ref{tab:zboxplots}. The boxes span the interquartile range with whiskers extending out to the full range of AUC values. For all three redshift scenarios we managed to reach our target purity of 95 per cent in three out of four algorithms. The relatively poor performance of ANN is attributed to the fact that these training samples are small compared to the test set, however, neural networks are known to perform well with large training samples (\citealt{Goodfellow-et-al-2016}, Section 1.2.2).

Fig. \ref{fig:representative-aucs} shows that, whilst there is overlap in the spread of AUC scores, the trend for all algorithms is an increase in mean and median, suggesting that redshift is a significantly impactful feature to the outcome of classification performance. The extent of improvement seems to be in agreement with the example of ROC curve results in \cite{2016ApJS..225...31L} (with the exception of KNN): AUC scores increase by -0.026, 0.016, 0.016 and 0.010 for KNN, SVM, ANN and BDT respectively. We see an increase in their average AUC scores of 0.003, 0.016, 0.020 and 0.012 (comparing No-z and Photo-z). The slight numerical discrepancies in AUCs may be due to splitting classification probabilities between the types Ia, Ibc and II, rather than just Ia and non-Ia as we have done here. Our finding that there is noticeable improvement when including redshift is in contrast to their conclusion that, when considering relative feature importance, redshift is fairly unimportant to this wavelet feature extraction method. 

We find similar results for both photometric and spectroscopic redshift, which may be explained by the absence of any catastrophic outliers in the simulated photometric redshifts in the SPCC; there is little scatter when comparing the two, with a root mean square error of only 0.028. This is perhaps optimistic, as it is estimated that around 10 per cent of galaxy photometric redshift results using LSST photometry will be outliers at $z=0.5$, reaching even higher percentages at lower redshifts (where outliers are those with redshift error greater than 3 times the robust standard deviation, or 0.06, as defined in \citealt{2020AJ....159..258G}).

Furthermore, the study \cite{2021PhRvD.103b3524M} finds that robust supernova cosmology cannot solely rely on photometric redshifts. Investigating the systematic requirements for a LSST-like survey, photometric redshifts at $z \lesssim 0.2$ in particular are found to be problematic, causing bias in dark energy cosmological inference. They conclude photometric redshifts should be used for cosmology only for $z>$~0.3 and spectroscopic follow-up should be conducted for all supernovae at $z\lesssim$~0.2--0.3.

In this comparison for representative samples we did not alter the photometric redshifts and we used them as they are in the SPCC. Irrespective of the use of either photometric or spectroscopic redshift as an additional feature for classification in this dataset, when the training sample is representative of the test set we observe promising results, including very high purity values. For the rest of this work we use spectroscopic redshifts, as discussed later in Section \ref{subsec:redshift}. Our aim is to at least match the classification performance that we would get when using a representative training sample, although, as we show in the next section, current 4MOST capabilities would only deliver a magnitude-limited sample. Our task is consequentially to improve upon a magnitude-limited sample to increase representativity. To address this, we add more training objects at fainter magnitudes and higher redshifts through two routes: observation with larger telescopes and augmentation.

\begin{table*}
    \centering
    \begin{tabular}{c|c|c|c|c|c|c|c}
        \hline
        Algorithm & Redshift & Mean & Median & IQR & Max & Min & Purity 95\% \\
        \hline
        \hline
        \multirow{3}{*}{KNN} & No-z & 0.959 & 0.958 & 0.003 & 0.964 & 0.953 & 20  \\
         & Photo-z & 0.962 & 0.963 & 0.003 & 0.969 & 0.956 & 20 \\
         & Spec-z & 0.964 & 0.965 & 0.004 & 0.970 & 0.959 & 20 \\
        \hline
        \multirow{3}{*}{SVM} & No-z & 0.947 & 0.949 & 0.008 & 0.955 & 0.929 & 18 \\
         & Photo-z & 0.963 & 0.964 & 0.004 & 0.967 & 0.950 & 11 \\
         & Spec-z & 0.963 & 0.965 & 0.005 & 0.969 & 0.949 & 12 \\
        \hline
        \multirow{3}{*}{ANN} & No-z & 0.914 & 0.913 & 0.008 & 0.926 & 0.906 & 0 \\
         & Photo-z & 0.934 & 0.936 & 0.007 & 0.943 & 0.920 & 1 \\
         & Spec-z & 0.934 & 0.935 & 0.006 & 0.942 & 0.924 & 1 \\
        \hline
        \multirow{3}{*}{BDT} & No-z & 0.953 & 0.954 & 0.004 & 0.959 & 0.938 & 20 \\
         & Photo-z & 0.965 & 0.964 & 0.005 & 0.973 & 0.958 & 20 \\
         & Spec-z & 0.966 & 0.967 & 0.007 & 0.971 & 0.959 & 20 \\
        \hline
    \end{tabular}
    \caption{AUC means, medians, interquartile ranges, maxima and minima for representative training samples over 20 runs, and the number of those runs that reached 95 per cent purity. These summarise the results shown in Fig. \ref{fig:representative-aucs} for the four different algorithms, comparing the cases of no redshift (No-z), photometric redshift (Photo-z) and spectroscopic redshift (Spec-z).}
    \label{tab:zboxplots}
\end{table*}

\newpage
\section{Simulating a 4MOST spectroscopic sample}
\label{sec:training}
\subsection{TiDES simulations}
\label{subsec:maglimit}
To determine the likely exposure times required for a TiDES sample of supernova spectra, we use a realistic mock catalogue containing supernovae with population fractions following the literature. Included supernova types are: Ia, split into normal Ia, 91T and 91bg using the fractions of each type given in \cite{2011MNRAS.412.1441L} and with a rate from \cite{2019MNRAS.486.2308F}; Core-collapse, split into Ib, Ic, IIL and IIP using the fractions given in \cite{2014AJ....147..118R}, with a rate proportional to the star-formation history in \cite{2008MNRAS.388.1487L}, anchored at low-redshift by the volumetric core-collapse rate from the Sloan Digital Sky Survey II Supernova Survey \citep{2014ApJ...792..135T}. The different supernova types and rates in the catalogue are necessary to reflect variations in spectra, which affect the rate of success in obtaining spectra of sufficient signal-to-noise ratio (SNR), defined later in this section. The LSST cadence assumed follows the \textsc{mothra\_2045} OpSim survey strategy\footnote{\url{https://www.lsst.org/scientists/simulations/opsim}}. The catalogue itself is limited at a peak magnitude of $r_{\textrm{AB}}$~=~24~mag, where any supernova that peaks brighter than this is simulated to be detected by LSST.

Each transient in the catalogue is assigned a spectrum from a set of templates based on its type, phase and magnitude. Additionally, for Type Ia supernovae, there is variation in their spectra based on the $x_0$, $x_1$ and $c$ SALT2 light curve parameters \citep{2007A&A...466...11G}. The spectra, normalised to the $r$-band magnitude at the time of observation, are run through the 4MOST exposure time calculator (ETC), which can quickly calculate exposure time requirements for thousands of targets. The ETC uses the 4MOST instrument response and outputs of the simulator TOAD (Top-Of-Atmosphere-to-Detector; \citealt{10.1117-12.2056463}), providing extensive modelling of both system throughput and sensitivity. The ETC is a parametrized version of TOAD, calculating the 1D signal and noisy spectra for targets with different target-fibre alignments and observing conditions such as sky brightness, transmission and seeing. By specifying a SNR (and given the magnitude of the targeted supernova), the ETC can return the target's required exposure time (and vice versa).

\subsection{Spectral success criterion}
\label{subsec:scc}
Our results come from running the catalogue through the 4MOST ETC v0.02 (in May 2019). However, since then, the ETC has been updated with newer versions. For a fixed exposure time and scaling results to the same effective SNR criterion, we find that the ETC v0.6 (in September 2020) agrees with the ETC v0.02 to within 0.02 mag, and so the difference was ignored. For TiDES supernovae, given a SSC, the success of observation depends upon both the spectral features present and the amount of `contaminating' light from the transient host galaxy \citep{2019Msngr.175...58S}. As supernova spectra are dominated by broad features, TiDES' SSC is defined using the average SNR over 15~\AA{} bins (over the range 4500-8000~\AA{} in the observed frame). TiDES' criterion is based on earlier studies of high-redshift Type Ia supernovae \citep{2009A&A...507...85B}, where robust classification is achieved with a mean SNR~=~5 per 15~\AA{} and probable classification of transients is demonstrated with a mean SNR as low as 3 per 15~\AA{}. However, in this study we adopt a more conservative criterion of SNR~=~7 per 15~\AA{}. We assume that all spectra that meet this criterion are correctly classified.

\begin{figure}
    \centering
    \includegraphics[width=\columnwidth]{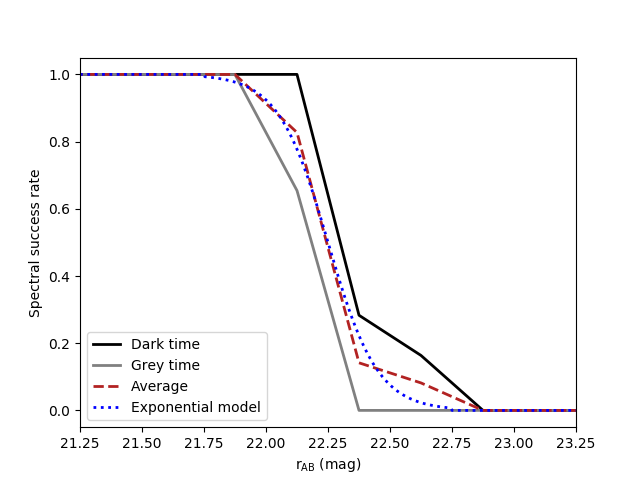}
    \caption{The success rate represents the probability that we obtain a spectrum of sufficient signal to noise, and therefore successful classification, of a targeted supernova of magnitude $r_{\textrm{AB}}$. The success rate is 50 per cent at $r_{\textrm{AB}}$~=~22.25~mag. The rate is calculated as the proportion of input supernovae for which a successful spectrum was obtained in magnitude bins of size 0.25. The average between the rates for dark and grey time is modelled as an exponential function (see Table \ref{tab:completeness}) that is later used to create our simulated training sample.}
    \label{fig:completeness}
\end{figure}

Current 4MOST simulations combine observing fields of the same sky coordinates and instrument position angle into observing blocks (OBs, \citealt{2020MNRAS.497.4626T}). The duration of the OBs are limited by a total exposure time of 1~h. Success is determined by whether a targeted supernova spectrum's necessary exposure time falls below 1~h for our criterion of SNR = 7 per 15~\AA{}. The rate of success for obtaining supernova classification from their spectra as a function of magnitude is shown in Fig. \ref{fig:completeness}. The success rate does not take into account 4MOST's fibre-target allocation which will depend on all 4MOST surveys and their science goals \citep{2020A&A...635A.101T}. Observation of each object in the catalogue was simulated in dark and grey time (we assume none of our targets will be targeted during bright time). Dark and grey time are defined by the amount of moon illumination (fraction of lunar illumination, FLI) where FLI$~<~$0.4 and 0.4$~\leq~$FLI$~\leq~$0.7 for dark and grey respectively\footnote{\url{https://www.eso.org/sci/observing/phase2/ObsConditions.html}}. The success rate is averaged over both curve distributions at each magnitude as the current dark/grey/bright cadence is undecided for 4MOST. However, current simulations for 4MOST's tiling pattern favour dark time over grey \citep{2020MNRAS.497.4626T}, so their average can be considered as a lower limit to our success rate. The function describing the success rate is shown in Table \ref{tab:completeness}. The exponential function in the second row was chosen to represent the average between dark- and grey-time success rates.

\begin{table}
    \centering
    \begin{tabular}{c|c}
         \hline
         Magnitude & Success rate \\
         \hline
         \hline
         $r_{\textrm{AB}}<21.75$ & 1.0 \\
         $21.75 \leq r_{\textrm{AB}} \leq 22.75$ & $\{1+\textrm{exp}[10 \; (r_{\textrm{AB}}-22.25)]\}^{-1}$ \\
         $r_{\textrm{AB}}>22.75$ & 0.0 \\
         \hline
    \end{tabular}
    \caption{The model for spectral success with 4MOST used to define the probability that an object observed with magnitude $r_\textrm{AB}$ will be selected for our simulated training sample.}
    \label{tab:completeness}
\end{table}

With 4MOST, it may be that we do not get the full 1~h observation for all our supernovae. This exposure time is based on two 30~min exposures in a single OB. Splitting into single exposures will affect the success rate of obtaining spectra of live transients. For the extreme (and unlikely) case in which all OBs contain only single 30~min exposures, the success rate curve keeps the same shape but moves $\sim$0.5~mag towards brighter magnitudes, i.e. 50 per cent success rate occurs at $r_{\textrm{AB}}$~=~21.75~mag. This would be much less favourable for our training sample prospects than for the success rate we simulate in Fig. \ref{fig:completeness}.

\subsection{Selecting training supernovae}
\label{subsec:selecting}
SPCC supernovae are selected for inclusion in the simulated training sample with a probability that follows the model described in Table \ref{tab:completeness}. However, to avoid using all supernovae brighter than the magnitude limit in the training and therefore not leaving any bright objects in the test set, the probabilities are scaled down by a factor of 2.

When selecting the training sample based on magnitude, the magnitudes used for each supernova are the $r_{\textrm{AB}}$ magnitudes closest to peak (i.e. the brightest $r$-band observation in the simulated light curve)\footnote{Previously, it was stated that the SPCC light curves consist of fluxes, however, each light curve point also has an assigned magnitude in the same band.}. By making the connection with Fig. \ref{fig:completeness}, which uses magnitude at the time of observation, we are assuming that we will obtain supernova spectra close to peak, within a few days. This is acceptable as, given 4MOST's limiting magnitude, we can only hope to get most objects' spectra close to peak.

\subsection{Use of redshift}
\label{subsec:redshift}
In general, 4MOST will not give us the opportunity to return to the same pointing of previously observed live transients and obtain a pure host-galaxy redshift. However, when we observe live transients, the light from the supernova and host galaxy will be blended and we expect to be able to measure host redshifts from these spectra, although not necessarily other host properties. This is what allows us to use the Type Ia in our spectroscopic sample for cosmology.

Science goal (ii) of TiDES will provide us with spectroscopic redshifts of many host galaxies of supernovae observed by LSST for which live spectroscopy was not possible. Hence, these are the transient objects that will define our test sample, i.e. the supernovae that we want to photometrically classify. We will therefore have a spectroscopic redshift for anything that makes it into our cosmological sample. As we will have spectroscopic redshift information for our training and test samples, in the following simulations in Section \ref{sec:results}, we use the spectroscopic redshifts of the SPCC simulated supernovae as an ancillary feature. This is the same as the spectroscopic redshift mentioned in Section \ref{subsec:representative}.

\section{Results}
\label{sec:results}

\begin{figure*}
    \begin{subfigure}[hb!]{.45\textwidth}
        \includegraphics[width=\textwidth]{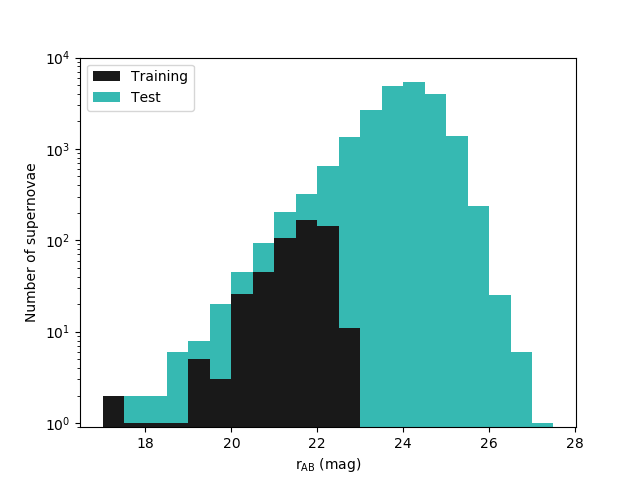}
        \caption{Stacked magnitude histogram of the training and test sets}
        \label{subfig:mag_hist_mag_lim}
    \end{subfigure}
    \begin{subfigure}[hb!]{.45\textwidth}
        \vspace{0.3cm}
        \includegraphics[width=\textwidth]{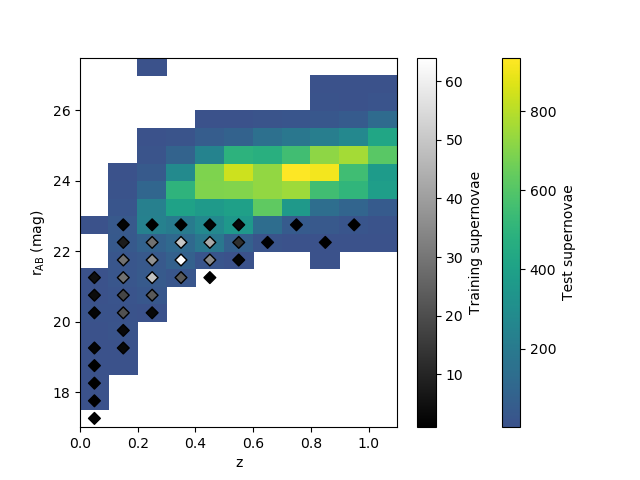}
        \caption{2D histogram of the relative distributions of redshift and magnitude in the training and test sets}
        \label{subfig:z-mag_mag_lim}
    \end{subfigure}
    \vskip\baselineskip        
    \begin{subfigure}[hb!]{0.45\textwidth}   
        \centering 
        \includegraphics[width=\textwidth]{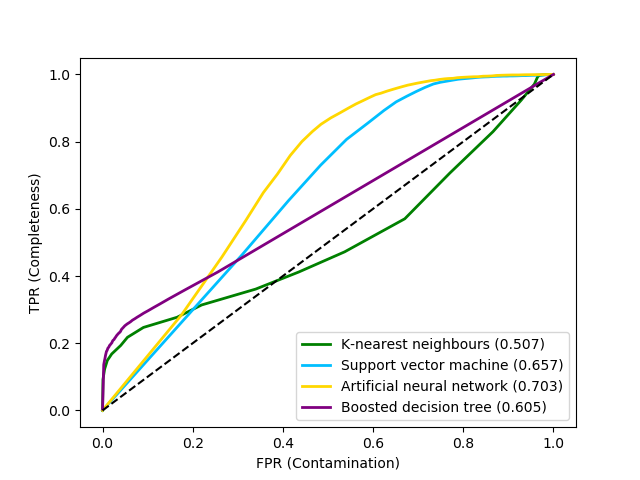}
        \caption{ROC curves}
        \label{subfig:roc_mag_lim}
    \end{subfigure}
    \quad
    \begin{subfigure}[hb!]{0.45\textwidth}  
        \centering 
        \includegraphics[width=\textwidth]{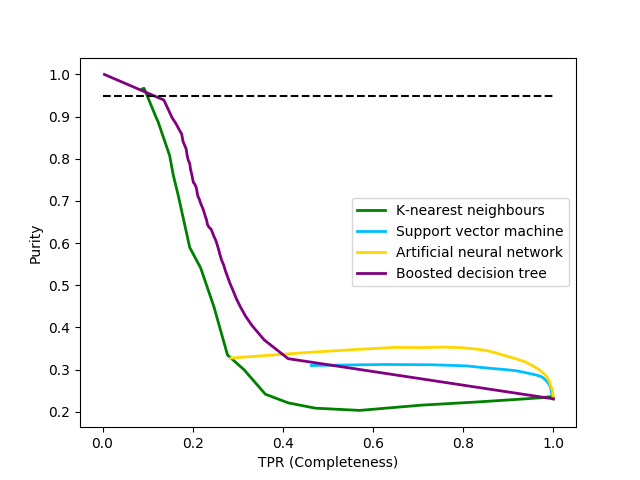}
        \caption{Purity curves}
        \label{subfig:tpr-com_mag_lim}
    \end{subfigure}
    \caption{Characteristics of the training and test samples for the case of a magnitude-limited training sample, and corresponding classification results are shown. This is one example from the 10 total runs. The ROC curves are close to resembling those of random classification. Despite the high purity reached for KNN (barely reaching the 95 per cent target) and BDT, the returned completeness of the classified sample is very low. Comparing to representative training, we are far from the classification algorithms' potential and need to improve upon this training sample. Our ROC curves pass through the two theoretical classification extremes: $(\textrm{TPR},\textrm{FPR}) = (0,0)$, in which everything is classified as non-Ia, and $(\textrm{TPR},\textrm{FPR}) = (1,1)$, in which everything is classified as Ia. It should be noted that our AUC scores are calculated using only TPR and FPR values from classification based on the used range of probability thresholds. If $\textrm{TP} = \textrm{FP} = 0$, then the purity is undefined. In these cases, the purity curve may not start at $\textrm{TPR} = 0$. This also occurs if the minimum TPR value from our range of probability thresholds is non-zero, as purity is undefined below this TPR.}
    \label{fig:mag_lim}
\end{figure*}

\begin{figure*}
    \begin{subfigure}{.45\textwidth}
        \includegraphics[width=\textwidth]{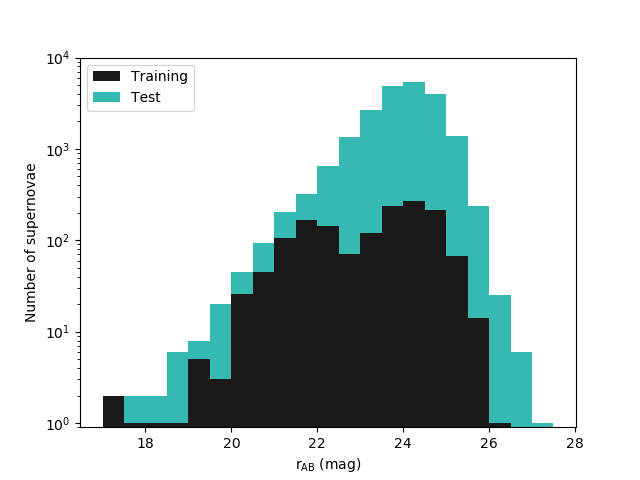}
        \caption{Stacked magnitude histogram of the training and test sets}
        \label{subfig:mag_hist_faint}
    \end{subfigure}
    \begin{subfigure}{.45\textwidth}
        \vspace{0.3cm}
        \includegraphics[width=\textwidth]{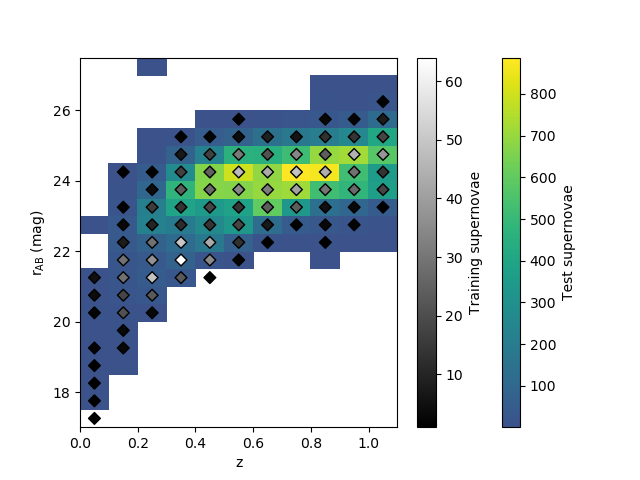}
        \caption{2D histogram of the relative distributions of redshift and magnitude in the training and test sets}
        \label{subfig:z-mag_faint}
    \end{subfigure}
    \vskip\baselineskip        \begin{subfigure}{0.45\textwidth}   
        \centering 
        \includegraphics[width=\textwidth]{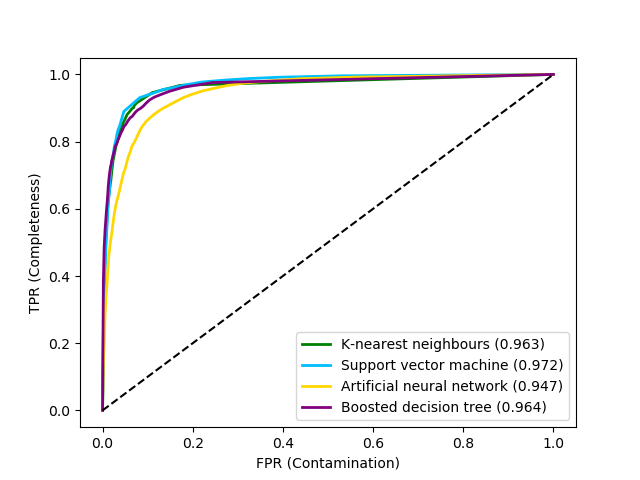}
        \caption{ROC curves}
        \label{subfig:faint_roc}
    \end{subfigure}
    \quad
    \begin{subfigure}{0.45\textwidth}  
        \centering 
        \includegraphics[width=\textwidth]{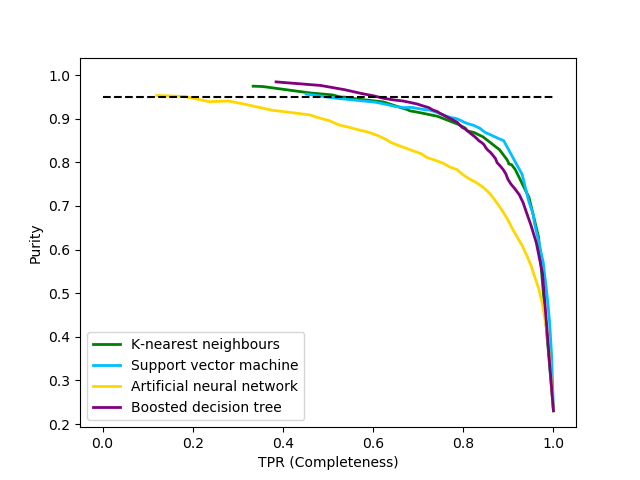}
        \caption{Purity curves}
        \label{subfig:tpr_com_faint}
    \end{subfigure}
    \caption{Characteristics of the training and test samples for the case of the previous magnitude-limited example (from Fig. \ref{fig:mag_lim}) but appended with fainter supernovae, and the corresponding results. Whilst we still don't have the same distributions of magnitude and redshift as the representative example in Fig. \ref{fig:representative}, the introduction of these objects into the training process has had a positive impact on the classification results.}
    \label{fig:mag_lim_faint}
\end{figure*}

\begin{figure*}
    \begin{subfigure}{.45\textwidth}
        \includegraphics[width=\textwidth]{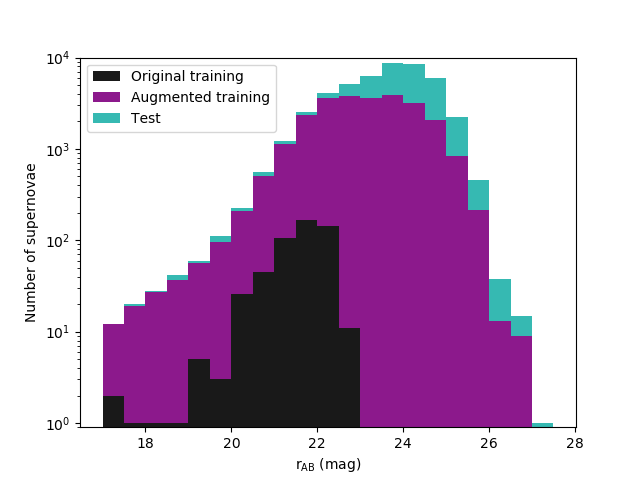}
        \caption{Stacked magnitude histogram of the original and augmented training, and test sets}
        \label{subfig:mag_hist_augmented}
    \end{subfigure}
    \begin{subfigure}{.45\textwidth}
        \vspace{0.3cm}
        \includegraphics[width=\textwidth]{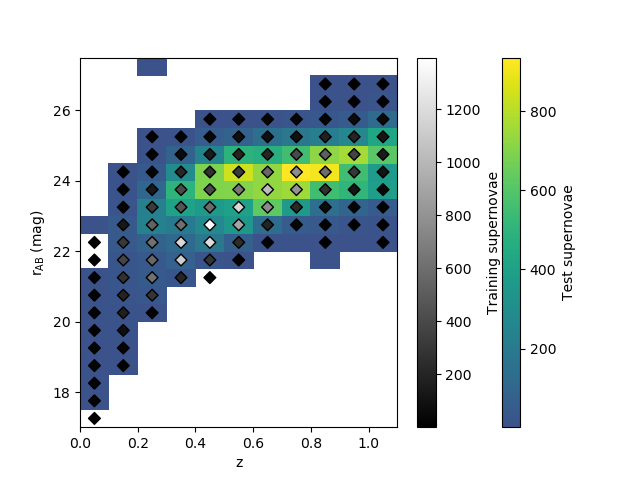}
        \caption{2D histogram of the relative distributions of redshift and magnitude in the augmented training and test sets}
        \label{subfig:z-mag_augmented}
    \end{subfigure}
    \vskip\baselineskip        
    \begin{subfigure}{0.45\textwidth}   
        \centering 
        \includegraphics[width=\textwidth]{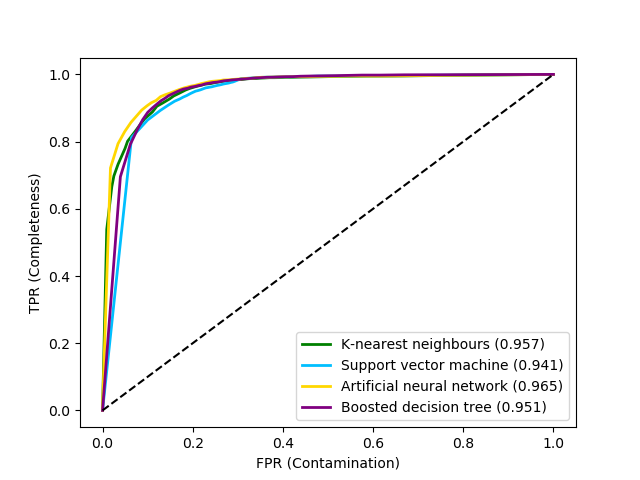}
        \caption{ROC curves}
        \label{subfig:roc_augmented}
    \end{subfigure}
    \quad
    \begin{subfigure}{0.45\textwidth}  
        \centering 
        \includegraphics[width=\textwidth]{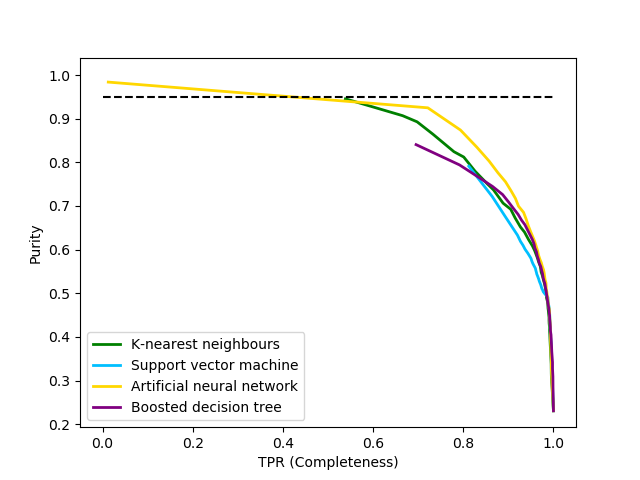}
        \caption{Purity curves}
        \label{subfig:tpr-com_augmented}
    \end{subfigure}
    \caption{We augment the original magnitude-limited training sample from Fig. \ref{fig:mag_lim}, increasing its size by a factor of 50 and extending it to much fainter magnitudes. The 2D histogram shows that the relative training and test numbers in each bin are not proportional, with more concentrated training supernovae at brighter magnitudes, although the training sample now covers the range of the test set.}
    \label{fig:augmented}
\end{figure*}

\begin{figure*}
    \begin{subfigure}{.45\textwidth}
        \includegraphics[width=\textwidth]{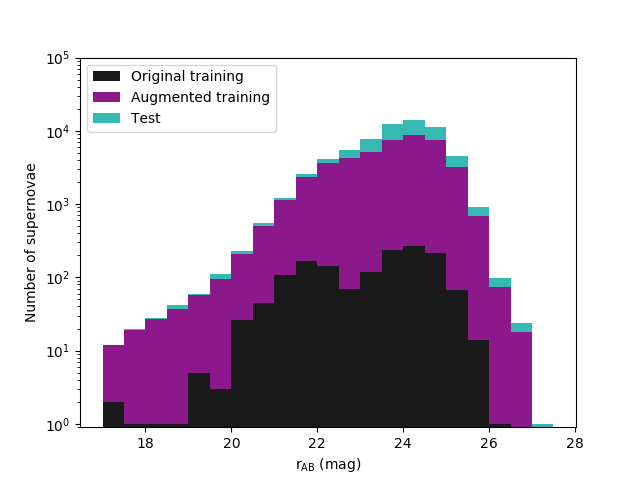}
        \caption{Stacked magnitude histogram of the original and augmented training, and test sets}
        \label{subfig:mag_hist_faint_augmented}
    \end{subfigure}
    \begin{subfigure}{.45\textwidth}
        \vspace{0.3cm}
        \includegraphics[width=\textwidth]{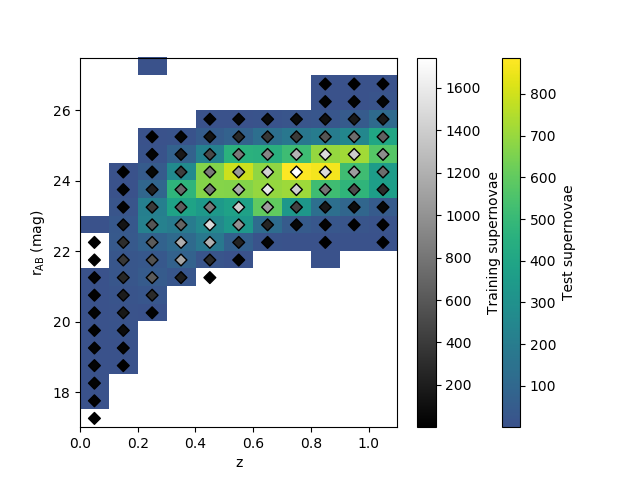}
        \caption{2D histogram of the relative distributions of redshift and magnitude in the augmented training and test sets}
        \label{subfig:z-mag_faint_augmented}
    \end{subfigure}
    \vskip\baselineskip        
    \begin{subfigure}{0.45\textwidth}   
        \centering 
        \includegraphics[width=\textwidth]{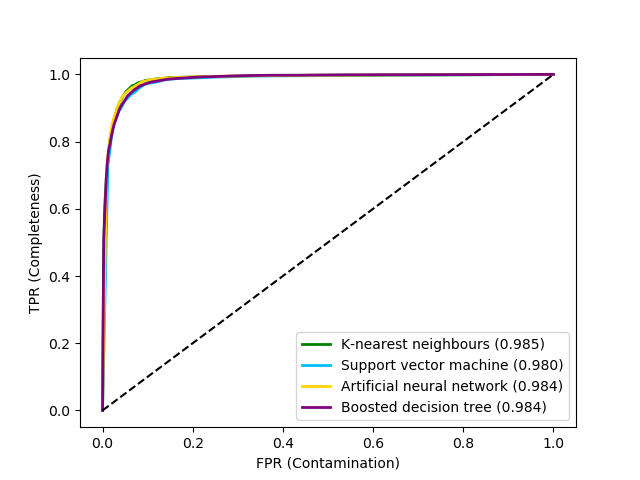}
        \caption{ROC curves}
        \label{subfig:roc_faint_augmented}
    \end{subfigure}
    \quad
    \begin{subfigure}{0.45\textwidth}  
        \centering 
        \includegraphics[width=\textwidth]{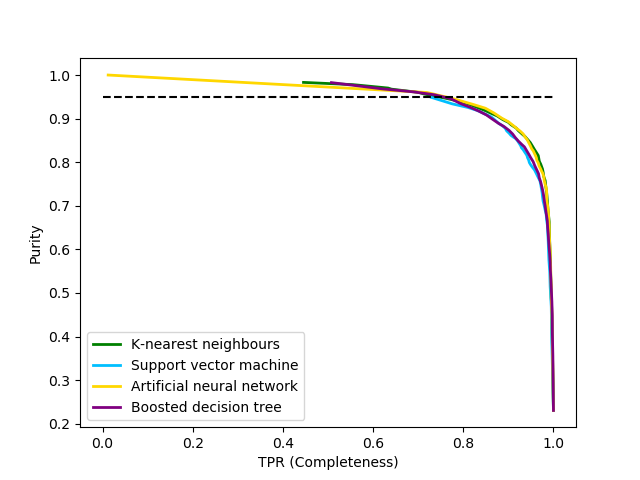}
        \caption{Purity curves}
        \label{subfig:tpr-com_faint_augmented}
    \end{subfigure}
    \caption{Our final main result is for the case of a magnitude-limited sample combined with additional faint supernovae that is then augmented. There is a higher concentration of fainter, high-redshift training supernovae than the previous augmented sample. This has improved further upon our results from Figs. \ref{fig:mag_lim_faint} and \ref{fig:augmented}.}
    \label{fig:add_faint_augmented}
\end{figure*}

Here we present the main results of the different classification simulations that we carry out, starting with the 4MOST magnitude-limited training sample. All our results are summarised in Table \ref{tab:boxplots} at the end of this section.

\subsection{A magnitude-limited training sample}
\label{subsec:mag_limited}
We first present our findings for a simulated training sample created following the method described in Section \ref{subsec:selecting}. We run this same test for 10 different training sets, where supernovae are sampled with probabilities following this spectral selection probability. This results in a magnitude-limited training sample of approximately 500 supernovae. Given that we are expecting a spectroscopic sample of size >30000 from TiDES, we would require a much larger dataset to fully simulate our prospective results. Nevertheless, by applying the 4MOST magnitude limit we are investigating its effect on algorithm-training and, ultimately, how to maximize our classification potential based on this observing constraint.

Fig. \ref{subfig:mag_hist_mag_lim} shows a stacked magnitude histogram of the training and test sets for one such magnitude-limited example. We show this alongside the distribution of training objects with respect to the test set (remaining objects from the SPCC) in redshift-magnitude space (Fig. \ref{subfig:z-mag_mag_lim}). Comparing to the representative training sample example in Figs. \ref{subfig:mag_hist_representative} and \ref{subfig:z-mag_representative}, clearly, a magnitude-limited training sample is not covering the full ranges of redshift and magnitude present in the test set. We examine the effect this has on the feature-space of the training supernovae with respect to the test set in Section \ref{subsec:discussion}.

Considering the ROC curves for this training sample (Fig. \ref{subfig:roc_mag_lim}), we find that the classifiers struggle to perform much better than random (shown by the dashed line) and are far from reaching the top-left corner. The magnitude limit has evidently had a negative impact on the classification.

We also find that it is difficult to reach high purities (Fig. \ref{subfig:tpr-com_mag_lim}). Often, it is impossible to reach a purity of 95 per cent and, even the few times it succeeded (generally requiring the maximum probability threshold), we return so few correct Type Ia supernovae that manipulating the classification parameters to achieve this would not significantly increase our cosmological sample. Also, not shown in the figure, we find that the completeness for high purities is consistently zero much beyond the faintest magnitudes of the training supernovae. For our purposes, the classification results are a failure when using a purely magnitude-limited training sample. For more practical uses, we instead therefore require methods to address this bias towards bright, low-redshift supernovae and produce a more representative training sample.

\subsubsection{Redshift in magnitude-limited training}
\label{subsubsec:z_results}
The magnitude limit seems to also imply a redshift limit (very few, if any, training supernovae are found beyond $z$~=~0.5--0.6, shown in Fig. \ref{subfig:z-mag_mag_lim}), although, depending on the specific sampling, the cut-off may not be as obvious. Inclusion of spectroscopic redshift, as opposed to none, in magnitude-limited training samples does not make a clear improvement to classification (comparing MagLim and MagLimNo-z in Table \ref{tab:boxplots}), as it did for representative training, discussed in Section \ref{subsec:representative}. A likely explanation of this is because including redshift in a magnitude-limited training sample does not give any extra information about fainter supernovae. Without redshift information, SVM and ANN perform worse and KNN and BDT seem to improve based on their mean and median scores. All four algorithms have a wide range of results, although they yield a higher maximum AUC when redshift is included.

\subsection{Reaching fainter magnitudes}
\label{subsec:fainter}
\subsubsection{Adding faint objects - making use of additional spectroscopic facilities}
\label{subsubsec:add-faint}
4MOST alone cannot provide us with a fully representative training sample. The required exposure times for supernovae fainter than $r_{\textrm{AB}} \approx 22.5$~mag are generally too large to consider spectroscopic follow-up with 4MOST. One option would be to use other spectroscopic facilities such as the VLT and ELT. Using the same SNR criterion as for 4MOST, we simulate a realistic faint spectroscopic sample of supernovae to combine with our 4MOST sample.

We simulate a total exposure time of 1000~h on the VLT and 100~h on the ELT (assuming 100~h and 10~h per 6-month semester over 5 years respectively). Individual supernova exposure times are determined using their brightest magnitudes and are based on calculations from the online ETCs. For the VLT ETC\footnote{\url{https://www.eso.org/observing/etc/bin/gen/form?INS.NAME=FORS+INS.MODE=spectro}} we use the FORS2 instrument and fix object (point source at $z = 0.6$) and sky parameters, varying the magnitude normalised in the $r$-band to estimate our exposure times. Parameters used are a moon FLI~=~0.2, airmass~=~1.50, seeing/image quality IQ~=~0.80~arcsec with a slit width of 1.00~arcsec using the GRIS\_300V+10 (>450nm,GG435) grism. Similarly, for the ELT ETC\footnote{\url{https://www.eso.org/observing/etc/bin/gen/form?INS.NAME=ELT+INS.MODE=swspectr}} we use airmass~=~1.50 and seeing~=~0.8~arcsec with the Laser-Tomography Adaptive Optics mode and radius of circular SNR ref. area~=~200~mas.

We randomly sample supernovae from the SPCC between $r_{\textrm{AB}}$~=~22.5--24.35 for the VLT and calculate their exposure times (adding an assumed overhead time of 5~min per object) until reaching the total. Supernovae with magnitudes of $r_{\textrm{AB}}$~>~24.35 would require $>$~4~h exposure time with the VLT. In our simulation these sources would be observed by the ELT. Hence, we similarly determine a ELT sample of supernovae with $r_{\textrm{AB}}>$~24.35 and 9~min overheads\footnote{6 min for guide star aquisition plus 3 min for the adaptive optics to produce the required image quality, described in the ELT Top Level Requirements at \url{https://www.eso.org/sci/facilities/eelt/docs/index.html}}. This produces a sample of $\sim$600 VLT supernovae and $\sim$400 ELT supernovae, increasing the total size of our simulated spectroscopic training sample by approximately 200 per cent ($\sim$500 to $\sim$1500). We acknowledge that in reality our magnitude-limited TiDES sample would be much larger, and hence our simulated TiDES sample is out of proportion to this realistic faint sample. A full accurately simulated sample is not possible with this dataset because of the relatively small number of bright objects.

With the addition of our faint sample we obtain the resulting magnitude and redshift-magnitude distributions in Figs. \ref{subfig:mag_hist_faint} and \ref{subfig:z-mag_faint}. We see a clear improvement on the overall performance of the machine-learning algorithms due to them making more informed classifications; the ROC curves (Fig. \ref{subfig:faint_roc}) have moved far away from the random classification associated with the diagonal dashed line. Going from the purely magnitude-limited training to the addition of fainter supernovae, over the 10 runs the average AUC increased from 0.547 to 0.961 for KNN, 0.671 to 0.960 for SVM, 0.702 to 0.946 for ANN and 0.628 to 0.969 for BDT.

Furthermore, we see significant improvements in the purity of our classified samples. Similar to the ROC curves reaching the top-left of the plot, good classification is also indicated by purity-completeness curves reaching the top-right, such as in the example in Fig. \ref{subfig:tpr_com_faint}. Notably, adding our faint sample into the training results in all 10 runs reaching 95 per cent purity for KNN and BDT (up from 2 and 7 respectively).

\subsubsection{Augmenting the training sample}
\label{subsubsec:augmented_results}
There is another avenue that can be taken to reach fainter magnitudes for our training sample. A fully representative spectroscopic training sample may not be necessary with the advent of data augmentation methods \citep{2018MNRAS.473.3969R, 2019AJ....158..257B}. In particular, \cite{2019AJ....158..257B} demonstrates that using expensive spectroscopic resources is not required when there are well-sampled, intermediate-redshift objects available for augmenting the training set. In our case the test set does not include any classes of objects that are not present in the training sample. If there are previously unforeseen objects in the test set that are not in the training sample, then augmentation cannot help. This was observed with class 99 in PLAsTiCC.

We adapt the source code, \textsc{avocado}, used in the winning solution to the PLAsTiCC challenge, to augment our magnitude-limited training sample by creating new artificial light curves that are resampled, shifted in time, and are at different redshifts for a range of observing conditions and uncertainties (our version is now included in the \textsc{avocado} GitHub)\footnote{\url{https://github.com/kboone/avocado}}. We use the same augmentation procedure of implementing a 2D Gaussian process (dimensions of time and wavelength), although we make certain changes to \textsc{avocado}, so that our augmented light curves are specific to our dataset and reflect the kinds of light curve that we want to classify. Firstly, we change the band central wavelengths to those of DES to match the SPCC light curves that we are using in the tests. These are used as wavelength coordinates in the Gaussian process. We also ensure that our augmented light curves have a number of observations consistent with SPCC. This is achieved by randomly sampling from a two-peaked distribution used to model the number of light curve observations in the original dataset. We use the same \textsc{avocado} constraints on augmented supernova redshifts to avoid the Gaussian process having to extrapolate far from the available data, where modelling uncertainties dominate its prediction ($0.95z_{\textrm{old}}<z_{\textrm{new}}<5z_{\textrm{old}}$ and $1+z_{\textrm{new}}<1.5(1+z_{\textrm{old}})$, explained fully in \citealt{2019AJ....158..257B}). The next part we change is the simulation of the light curve uncertainties. As with the original method in \textsc{avocado}, all the SPCC's error bars in each band are well-modelled as lognormal distributions and so we use the lognormal parameters for our dataset's band noises to sample flux errors and set the depth of observations in our new light curves. Finally, we implement a method to check whether a new light curve would be useful in the context of our dataset and simulations. The pass criterion is that the new light curve contains simulated observations in the $r$-band, including a positive maximum flux (used to give the supernova's reference magnitude). Additionally, we discard any augmented light curves that have redshifts and magnitudes that fall outside the ranges in the SPCC. We do not have need of the original \textsc{avocado} methods of preprocessing light curves (accounting for consistent background flux levels) or augmenting galactic objects (objects in the PLAsTiCC dataset that have $z=0$).

For augmenting our magnitude-limited training sample, we use the 2D Gaussian process method in \textsc{avocado} to create up to 50 new versions of each original training supernova. We find that 50 is sufficient by augmenting our magnitude-limited training sample in multiples of 10 from 10 to 100 where classification (AUC) plateaus for around 40--50 new objects per original light curve. We do not reuse the same augmented light curves, but instead create a new set of augmented light curves for each run.

As augmentation simulates new objects at different redshifts, it therefore requires initial cosmological assumptions\footnote{This is done using \textsc{astropy.cosmology.FlatLambdaCDM} with Hubble parameter $H_0 =$~70~kms$^{-1}$Mpc$^{-1}$ and matter density parameter $\Omega_{\textrm{M}} = 0.3$}. Before using such a method in a real cosmological analysis, it will be important to test (with simulations) the impact of these assumptions on the final cosmological results. This is planned for a future investigation.

The first augmented training samples we create are from our magnitude-limited samples discussed in Section \ref{subsec:maglimit}. For these we augment the training to extend to fainter magnitudes and higher redshifts as shown in Figs. \ref{subfig:mag_hist_augmented} and \ref{subfig:z-mag_augmented}.
Without using any of the original SPCC supernovae beyond $r_{\textrm{AB}} \approx 22.5$~mag, augmentation of the training sample has introduced the algorithms to the features associated with faint light curves. Comparing these results (Figs. \ref{subfig:roc_augmented} and \ref{subfig:tpr-com_augmented}) to those of previous training samples, we again see a significant improvement over the magnitude-limited training sample (Figs. \ref{subfig:roc_mag_lim} and \ref{subfig:tpr-com_mag_lim}). However, compared to the magnitude-limited plus faint training sample (Figs. \ref{subfig:faint_roc} and \ref{subfig:tpr_com_faint}), despite being much larger we do not reach the same classification performance. This is with the exception of ANN, which, as expected, does well when presented with large training samples (\citealt{Goodfellow-et-al-2016}, Section 1.2.2), achieving higher AUC scores and more runs that reach 95 per cent purity.

\begin{figure*}
    \includegraphics[width=.92\textwidth]{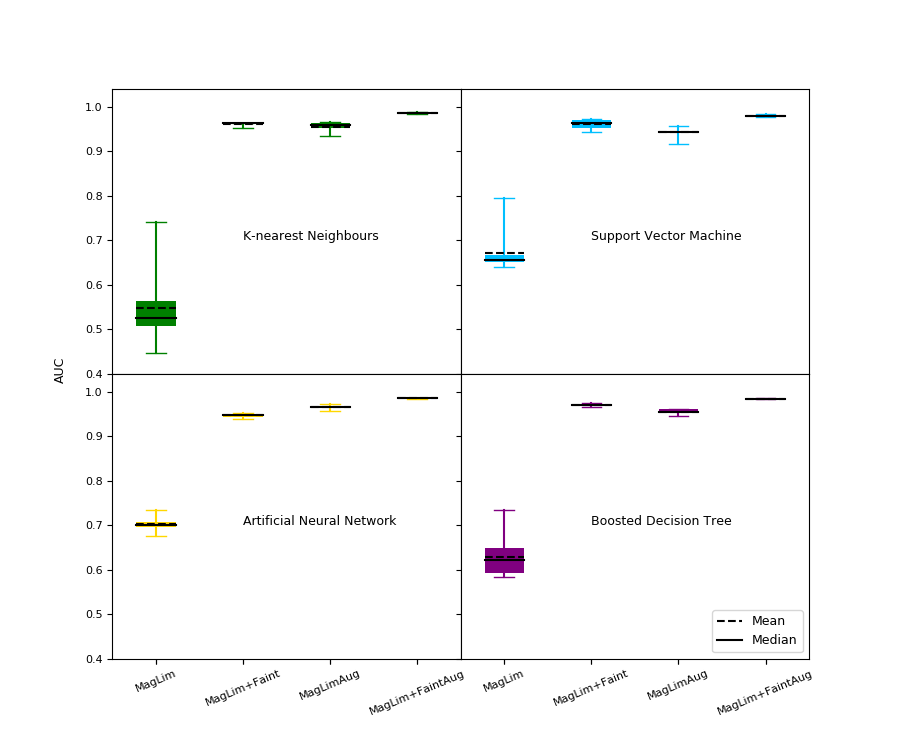}
    \caption{Boxplots showing the AUC scores over 10 runs for each of the four algorithms in four of our training sample simulations. The boxes represent the interquartile ranges, with their values shown in Table \ref{tab:boxplots}, along with means and medians. These results are for binary classification from Ia vs. non-Ia class probabilities. They are defined as MagLim: magnitude-limited training sample; MagLim+Faint: magnitude-limited sample with additional fainter superovae; MagLimAug: magnitude-limited sample augmented; MagLim+FaintAug: combined magnitude-limited and faint samples both augmented.}
    \label{fig:aucs}
\end{figure*}

Going one step further, we augment the combined magnitude-limited and faint supernovae sample (from Section \ref{subsubsec:add-faint}), shown in Figs. \ref{subfig:mag_hist_faint_augmented} and \ref{subfig:z-mag_faint_augmented}. This differs from the previous augmented training sample as we now start with `true' supernova light curves from fainter magnitudes, enabling the augmentation procedure to create more realistic faint light curves. The introduction of these produces our most successfully classified samples and is seen by the trend in the AUC boxplots in Fig. \ref{fig:aucs} and is summarised in Table \ref{tab:boxplots}.

\begin{figure*}
    \begin{subfigure}{.45\textwidth}
        \includegraphics[width=\textwidth]{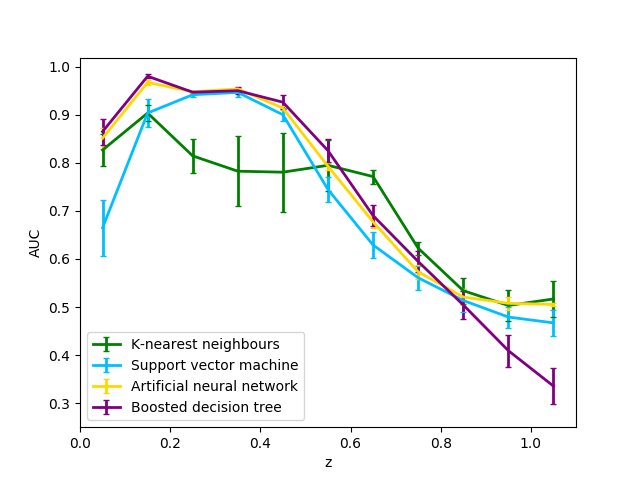}
        \caption{Magnitude-limited training sample}
        \label{subfig:auc-z-mag-lim}
    \end{subfigure}
    \begin{subfigure}{.45\textwidth}
        \vspace{0.25cm}
        \includegraphics[width=\textwidth]{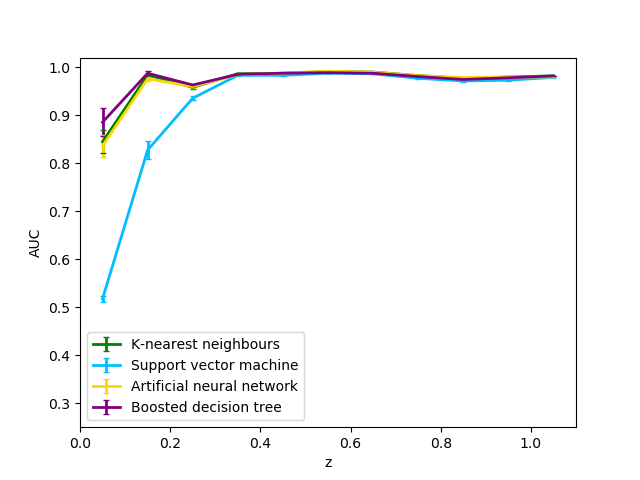}
        \caption{Augmented faint-appended magnitude-limited training sample}
        \label{subfig:auc-z-add-faint-augmented}
    \end{subfigure}
    \caption{Average AUC scores as a function of redshift for all four algorithms, calculated in bins of size 0.1. Error bars represent the standard error in the average over the 10 runs.}
    \label{fig:auc-z}
\end{figure*}

\begin{figure*}
    \includegraphics[width=\textwidth]{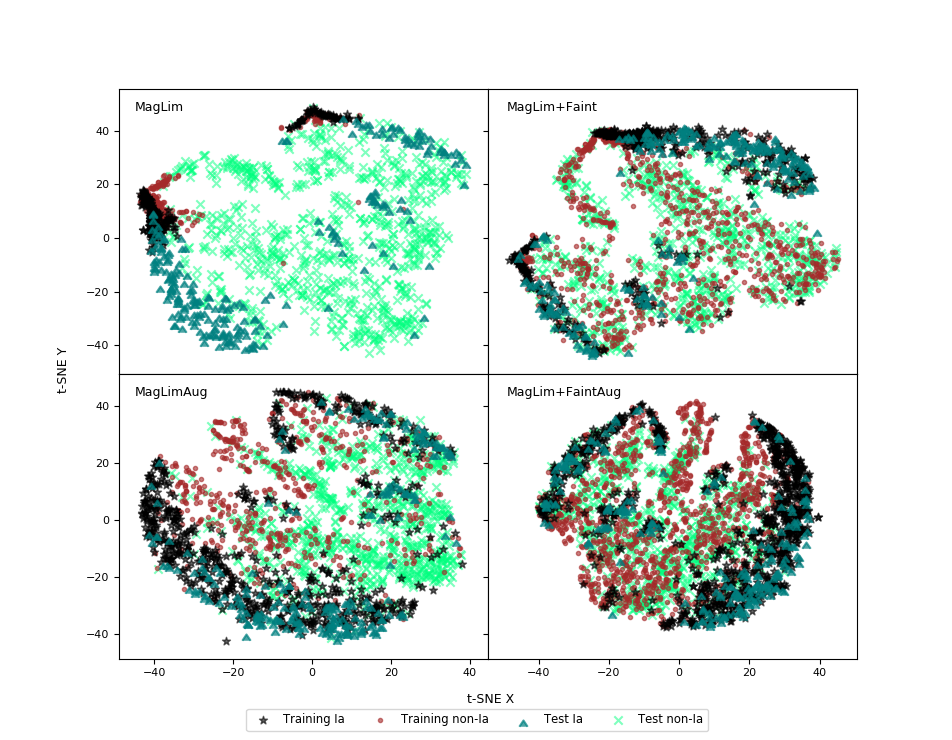}
    \caption{t-SNE plots comparing the feature-space coverage of four of our training samples. For clarity in these plots, one twentieth of the test set is shown for all, while one twentieth of the training is also shown for the augmented cases.}
    \label{fig:tsne-compare}
\end{figure*}

\begin{table*}
    \centering
    \begin{tabular}{c|c|c|c|c|c|c|c}
        \hline
        Algorithm & Training & Mean & Median & IQR & Max & Min & Purity 95\% \\
        \hline
        \hline
        \multirow{9}{*}{KNN} & MagLim & 0.547 & 0.525 & 0.056 & 0.741 & 0.446 & 2 \\
         & MagLim+Faint & 0.961 & 0.962 & 0.001 & 0.964 & 0.953 & 10 \\
         & MagLimAug & 0.955 & 0.958 & 0.011 & 0.964 & 0.934 & 3 \\
         & \textbf{MagLim+FaintAug} & \textbf{0.985} & \textbf{0.985} & \textbf{0.001} & \textbf{0.987} & \textbf{0.983} & \textbf{10} \\
         & FaintAug & 0.973 & 0.974 & 0.004 & 0.980 & 0.958 & 10 \\
         & MagLimNo-z & 0.631 & 0.640 & 0.053 & 0.707 & 0.536 & 0 \\
         & MagLim3Class & 0.543 & 0.525 & 0.064 & 0.711 & 0.436 & 2 \\
         & \textbf{MagLim+FaintAug3Class} & \textbf{0.985} & \textbf{0.985} & \textbf{0.001} & \textbf{0.987} & \textbf{0.983} & \textbf{10} \\
         & MagLimAug+Faint & 0.976 & 0.976 & 0.001 & 0.978 & 0.972 & 10 \\
         \hline
         \multirow{9}{*}{SVM} & MagLim & 0.671 & 0.657 & 0.015 & 0.795 & 0.640 & 1 \\
         & MagLim+Faint & 0.960 & 0.963 & 0.018 & 0.972 & 0.942 & 4 \\
         & MagLimAug & 0.942 & 0.944 & 0.005 & 0.956 & 0.916 & 3 \\
         & MagLim+FaintAug & 0.980 & 0.980 & 0.002 & 0.984 & 0.976 & 3 \\
         & FaintAug & 0.953 & 0.960 & 0.009 & 0.965 & 0.914 & 6 \\
         & MagLimNo-z & 0.617 & 0.601 & 0.088 & 0.714 & 0.504 & 2 \\
         & MagLim3Class & 0.670 & 0.660 & 0.019 & 0.748 & 0.646 & 2 \\
         & \textbf{MagLim+FaintAug3Class} & \textbf{0.982} & \textbf{0.983} & \textbf{0.001} & \textbf{0.986} & \textbf{0.976} & \textbf{10} \\
         & MagLimAug+Faint & 0.968 & 0.969 & 0.004 & 0.972 & 0.958 & 1 \\
         \hline
         \multirow{9}{*}{ANN} & MagLim & 0.702 & 0.700 & 0.011 & 0.734 & 0.675 & 0 \\
         & MagLim+Faint & 0.946 & 0.948 & 0.005 & 0.952 & 0.939 & 6 \\
         & MagLimAug & 0.965 & 0.964 & 0.005 & 0.973 & 0.956 & 9 \\
         & MagLim+FaintAug & 0.985 & 0.985 & 0.005 & 0.986 & 0.984 & 10 \\
         & FaintAug & 0.957 & 0.962 & 0.027 & 0.978 & 0.926 & 9 \\
         & MagLimNo-z & 0.613 & 0.614 & 0.060 & 0.688 & 0.568 & 0 \\
         & MagLim3Class & 0.711 & 0.708 & 0.042 & 0.803 & 0.621 & 0 \\
         & \textbf{MagLim+FaintAug3Class} & \textbf{0.986} & \textbf{0.986} & \textbf{0.001} & \textbf{0.988} & \textbf{0.984} & \textbf{10} \\
         & MagLimAug+Faint & 0.978 & 0.977 & 0.002 & 0.982 & 0.976 & 9 \\
         \hline
         \multirow{9}{*}{BDT} & MagLim & 0.628 & 0.622 & 0.057 & 0.733 & 0.584 & 7 \\
         & MagLim+Faint & 0.969 & 0.969 & 0.002 & 0.974 & 0.964 & 10 \\
         & MagLimAug & 0.955 & 0.954 & 0.008 & 0.960 & 0.946 & 1 \\
         & \textbf{MagLim+FaintAug} & \textbf{0.984} & \textbf{0.984} & \textbf{0.001} & \textbf{0.985} & \textbf{0.983} & \textbf{10} \\
         & FaintAug & 0.956 & 0.959 & 0.007 & 0.974 & 0.922 & 10 \\
         & MagLimNo-z & 0.642 & 0.635 & 0.062 & 0.709 & 0.560 & 6 \\
         & MagLim3Class & 0.645 & 0.650 & 0.081 & 0.710 & 0.580 & 6 \\
         & MagLim+FaintAug3Class & 0.979 & 0.980 & 0.002 & 0.981 & 0.977 & 4 \\
         & MagLimAug+Faint & 0.976 & 0.976 & 0.001 & 0.978 & 0.974 & 4 \\
         \hline
    \end{tabular}
    \caption{AUC means, medians, interquartile ranges, maxima and minima for different types of training sample over 10 runs, and the number of those runs that reached 95 per cent purity. The first four rows for each algorithm are the results shown in Fig. \ref{fig:aucs}. We compare our results with additional training samples, including just the augmented faint sample (FaintAug). We investigate how \textsc{snmachine} performs when returning 3 class probabilities (Ia, Ibc and II) for each supernova in the test set for magnitude-limited and augmented magnitude-limited-plus-faint samples (MagLim3Class and MagLimFaintAug3Class). Also, for the magnitude-limited case, we include results when using no redshift (MagLimNo-z). Finally, we include results from the runs investigating how adding the fainter supernovae (not augmented) on to the augmented magnitude-limited sample affected results (MagLimAug+Faint). We highlight in bold our most successful training sample for each algorithm, which is either MagLim+FaintAug, or MagLim+FaintAug3Class.}
    \label{tab:boxplots}
\end{table*}

\subsection{Ia vs. Ibc vs. II classification}
\label{subsec:3class}
We also run tests in which the \textsc{snmachine} algorithms are trained to recognize supernovae as being either Type Ia, Ibc or II, rather than the baseline Ia vs. non-Ia. This is done for the original magnitude-limited samples and the augmented magnitude-limited plus faint sample (comparing MagLim and MagLim+FaintAug to MagLim3Class and MagLim+FaintAug3Class respectively in Table \ref{tab:boxplots}). Considering mean AUC scores for MagLim, there is a small increase for ANN and BDT although no significant difference observed by making this change to classification.

For the augmented case, AUC scores are mostly very similar, the biggest change being a drop in average AUC of 0.05 for BDT. Also, quite notably for BDT, having 3 classes causes fewer runs to reach 95 per cent purity - decreasing from 10 to 4. Conversely, SVM sees an increase from 3 to 10 runs whilst keeping AUC scores fairly consistent. In this 3-class scenario for our most successful type of training sample, it appears that SVM would be a better choice than BDT, although BDT would be more suited in all other cases that we tested. ANN also performs better with 3 classes, although this change is negligible; the AUC scores barely change at the third significant figure. No change at all is seen for KNN.

\subsection{Discussion}
\label{subsec:discussion}
In Fig. \ref{fig:auc-z} we show the average AUC-dependence on redshift, comparing results for the original magnitude-limited training sample with the final augmented training sample (MagLim and MagLim+FaintAug). The large increase in size of training sample when augmenting has likely contributed to the effect of more predictable behaviour in the algorithms, shown by the very small error bars. Not only has augmented training improved the AUC scores at high-redshift, but also generally in low-redshift regions already covered by the magnitude-limited training sample. However, while consistently close to AUC~=~1 at high redshift, further improvement is required for $z<0.3$. Interestingly, the 0.0--0.1 and 0.1--0.2 bins for SVM actually performed worse for our most successful training sample.

We show t-SNE plots comparing training samples (MagLim, MagLim+Faint, MagLimAug, MagLim+FaintAug) in Fig. \ref{fig:tsne-compare} to help visualise why greater success is found with the addition of faint supernovae and augmentation. In each panel, the test set is the same, but the feature-space covered by the training sample changes. For magnitude-limited training, only a small region of the test set feature-space is covered for both Type Ia and non-Ia supernovae, demonstrating why classification performance is not very successful. When we consider the proportions of different supernova types in magnitude-limited training (see appendix \ref{app:proportions}, Fig. \ref{subfig:fracs-mag-lim}), we find that the bias towards bright objects is also a bias towards Type Ia supernovae. Evidently, this is why there are so few training non-Ia supernovae in the feature-space occupied by those in the test set. However, when the training sample is appended with faint supernovae and augmented, the training sample itself is not only much larger, but a significantly larger proportion of the feature-space is now covered for the respective supernovae types, similar to the case for a representative sample as shown in Fig. \ref{fig:tsne-representative}. For MagLimAug, there are still some regions of feature-space that are not covered, showing why this training sample does not perform as well as either MagLim+Faint or MagLim+FaintAug. This is likely due to \textsc{avocado}'s redshift constraints, preventing extrapolation far from where there is available data. It should be noted that the fractions of the supernova types in the augmented training sample are the same as those present in the original magnitude-limited sample, meaning that there remains a larger proportion of Type Ia supernovae in the training sample than in the test set (Fig. \ref{subfig:fracs-mag-lim-aug}; we briefly address this issue of balance in Appendix \ref{app:proportions}).

Augmentation enables us to fill in some of the significant gaps in the test set feature-space that we may not fully cover with our spectroscopic sample. Even though we have `true' faint supernovae to help train the algorithms with improved representativity, when combining with the augmented magnitude-limited sample it is better to augment these as well, as shown in the results summary table (Table \ref{tab:boxplots}). Fig. \ref{fig:aucs} shows that we need these `true' faint supernovae in our spectroscopic sample to achieve the highest AUC scores. There is a clear improvement in all algorithms going from purely magnitude-limited (MagLim) to adding faint supernovae (MagLim+Faint). As previously stated, the same positive trend is seen when these training samples are augmented (MagLimAug to MagLim+FaintAug).

We also compare these results to the hypothetical case of only having the faint sample and then augmenting that, with its results summarised in Table \ref{tab:boxplots} as FaintAug. When we augment just the faint sample of supernovae, we get AUC scores very similar to those for MagLim+Faint (and MagLimAug). This seems to indicate that the original magnitude-limited sample may not be so crucial for training, as similar success is found by augmenting just a faint spectroscopic sample, however, we fundamentally do also need our magnitude-limited TiDES sample to obtain the best classification results. Furthermore, classification performance in the case of the augmented faint sample suffers due to the similar coverage issue of the original magnitude-limited sample, but at the other end of the brightness scale.

As an attempt to save on computing time and resources, we also consider the MagLimAug+Faint training, i.e. augmenting just the magnitude-limited sample and then adding the non-augmented faint sample of supernovae. However, compared to MagLim+FaintAug(3Class), this is not as favourable for classification in terms of AUC or purity.

With an original spectroscopic sample extending as faint as possible, these results highlight the important role of augmentation to achieve successful photometric classification in future supernova surveys. For this particular purpose, KNN, SVM, ANN and BDT all appear to be reliable machine-learning algorithms, reaching high AUC scores with very small variations, and also being able to achieve 95 per cent purity over all test runs.

\section{Conclusions}
\label{sec:conclusions}
4MOST-TiDES expects to obtain the largest spectroscopically confirmed sample of supernovae to date (>30000), including Type Ia supernovae which will be used for precision cosmology. However, the transients that are not followed up spectroscopically may still be useful for cosmology. Herein lies the necessity for photometric classification. Using the capabilities and survey constraints of 4MOST, we forecast a spectroscopic sample of supernovae that is magnitude-limited, reaching $r_{\textrm{AB}} \approx$ 22.5~mag. Using machine-learning algorithms, we find the greatest success in the results of photometric classification when we combine this sample with fainter supernovae obtained from larger spectroscopic facilities and then augment the whole sample, to be used as a training set. Whilst on its own, 4MOST cannot give us a fully representative training sample, the accumulated dataset will provide an important basis for a training sample to photometrically classify other LSST transients for which we have host-galaxy redshifts. Including our photometrically classified sample, we expect to produce the largest ever cosmological sample of Type Ia supernovae by more than an order of magnitude.

In this paper, we started by demonstrating that a representative training sample (of size 1103) will yield good classification results with \textsc{snmachine}: AUC~>~0.9 and consistently high purities reaching 95 per cent (with the exception of ANN, although it is important to note that ANN will outperform the other algorithms with much larger training samples). This success is attributed to the fact that the algorithms are trained on features associated with the full range of magnitudes and redshifts in the test set. However, we find that a representative training sample of this nature will not be easily attainable with present spectroscopic facilities. These tests using representative training were also carried out to investigate the role of redshift as an additional feature for classification. We find a consistent improvement in AUC scores when including redshift, demonstrated by a noticeable increase in mean and median over 20 runs. Our results are similar to those in \cite{2016ApJS..225...31L}, although we consider inclusion of redshift important due to its significant impact on classification performance, in contrast to their conclusion that redshift is a relatively unimportant feature. Going from no redshift to photometric and spectroscopic redshifts respectively, we get an increase in average AUC over 20 runs from 0.959 to 0.962 and 0.964 for KNN, 0.947 to 0.963 (both redshifts) for SVM, 0.914 to 0.934 (both redshifts) for ANN and 0.953 to 0.965 and 0.966 for BDT. There appears to be no clear winner between photometric or spectroscopic redshift for this particular simulated dataset, both achieving very similar results. This is surprising, given the fact that photometric redshifts are usually less accurate and less precise than spectroscopic redshifts. We attribute the result to the minimal scatter between spectroscopic and photometric redshifts in the SPCC; the root mean squared error in photometric redshifts is very small (0.028). However, we find that when the training sample is magnitude-limited, it is less clear whether having redshift helps in the training process or not.

Whilst being a reliable dataset central to a number of previous supernova classification studies, the SPCC is not large enough that we can fully simulate a 4MOST spectroscopic sample. We find that when considering a spectroscopic sample that is magnitude-limited based on our success criteria and considering 4MOST's capabilities, there are so few objects in the SPCC (approximately 500 after scaling down by a factor of 2, out of 21319 in total, as discussed in Section \ref{subsec:maglimit}; we are only simulating approximately 1.6 per cent of the full TiDES sample) that our results are sensitive to specific choices of which supernovae we include in our training. Despite the variation and spread of results, it is clear that a magnitude limit implies a non-representative training sample that has poor coverage of the test-set feature-space (we show this in a t-SNE plot), and, therefore, very negatively affects our results. This does mean, however, that any significant improvement to the performance of the \textsc{snmachine} algorithms when dealing with magnitude-limited training samples is promising. The full TiDES sample size may improve the performance of the magnitude-limited training somewhat, but it will still suffer from the lack of coverage at faint magnitudes and high redshifts. The performance of the full TiDES training sample will be investigated in future work using a significantly larger simulation.

With our 4MOST magnitude-limited training sample as a basis, we next investigate how our results change when combining with additional faint supernovae. A realistic scenario for following up LSST alongside 4MOST would be obtaining spectra of fainter supernovae using facilities such as the VLT and ELT. We simulate such a scenario with the dataset, extending the training to high redshift and increasing the sample size by $\sim$1000 supernovae. Over 10 runs we see an increase in the average AUC from 0.547 to 0.961 for KNN, 0.671 to 0.960 for SVM, 0.702 to 0.946 for ANN and 0.628 to 0.969 for BDT. In particular, both KNN and BDT achieved a classified sample of 95 per cent purity in all 10 runs. This is a substantial boost to our results from our orginal training sample, although, on its own, it is not the most successful that we tested. Our results show that complementary  faint objects can significantly improve upon a 4MOST magnitude-limited training sample.

We next consider data augmentation to investigate further improvement. By creating artificial light curves, the size is limited only computationally, although we find that results plateau around 40--50 per original supernova. Applying \textsc{avocado} to the SPCC, we increase training sample size by a factor of 50. For the augmented magnitude-limited sample we reach average AUC scores of 0.955 for KNN, 0.942 for SVM, 0.965 for ANN and 0.955 for BDT; a large increase, but, with the exception of ANN, is not as successful as our combined magnitude-limited and faint training sample. When we augment the combined magnitude-limited and faint sample, we achieve our best AUC scores. However, there is a slight dependence on whether we train our machine learning algorithms to recognise supernovae as either Type Ia or non-Ia, or Type Ia, Ibc or II. Highest average AUCs are 0.985 for KNN, 0.982 for SVM, 0.986 for ANN and 0.984 for BDT and all algorithms are able to reach 95 per cent purity in all 10 runs for this training sample. Considering three classes appears most beneficial for SVM, as this is the only type of training we tested that resulted in all 10 runs reaching 95 per cent purity for this algorithm. For BDT, two classes is more favourable, and for KNN and ANN there is little to no difference. We attribute this success to the fact that including fainter supernovae adds some real constraints to the wavebands at faint magnitudes, i.e. \textsc{avocado} does not need to purely extrapolate from a set of bright, low-redshift supernovae, as it did when augmenting a purely magnitude-limited sample.

TiDES plans to blanket target every possible transient that is brighter than $r_{\textrm{AB}}$~=~22.5 mag. This will also avoid creating an artificially biased sample. In this work we assume that we have the full 4MOST-TiDES spectroscopic sample as a training data basis. Hence, our focus on optimisation is how to improve classification using this sample. However, there may be room for further optimisation in survey strategy in how we decide which transients to target that are just below this magnitude limit. Initially, we consider classification when using a hypothetical representative sample, although this may not reflect a fully optimised training sample. A fully optimised sample may require relatively overpopulated bins at high and low redshifts when compared to a `representative' sample. Achieving this in a spectroscopic follow-up survey would likely need to make use of active learning, following such methods as those presented in \cite{2019MNRAS.483....2I}.

Starting with a magnitude-limited training sample constrained by the capabilities of 4MOST, we find that it is optimised when combined with complementary faint supernovae and then augmented to have more coverage of the corresponding test set feature-space. Augmentation is a necessary step to create the most successful realistic training samples, although in future work it will be necessary to test how cosmological assumptions for augmentation could be creating potential bias. Furthermore, in our simulations we assume that the classifications in our spectroscopic sample are 100 per cent correct. Hence, we would want to investigate whether mis-classification of a 4MOST spectrum could propagate through the machine-learning pipeline and affect results, and ultimately the resultant cosmology we determine using our classified sample. These tests would ideally be done with a much larger dataset of supernovae to better reflect what we can do in reality.

\section*{Acknowledgements}
JEC acknowledges support from a STFC Data Science studentship and funding of training through the STFC 4IR Centre for Doctoral Training.
IMH acknowledges support for this work from STFC (consolidated grant numbers ST/R000514/1 and ST/P00038X/1).
AGK is supported by the U.S.\ Department of Energy, Office of Science, Office of High Energy 
Physics, under contract No.\ DE-AC02-05CH11231.
We thank Emille E. O. Ishida for useful discussions and for her very helpful comments in preparing this paper.
We also thank the reviewer at MNRAS for providing helpful and detailed comments that have significantly improved this paper.
We thank Lancaster University's High End Computing Cluster service for providing access to conduct our high performance computing needs in this work.
This work would not have been possible without the software \textsc{snmachine}, and so we also thank the developers Michelle Lochner, Jason D. McEwen and Hiranya Peiris.
The DESC acknowledges ongoing support from the Institut National de Physique Nucl\'eaire et de Physique des Particules in France; the Science \& Technology Facilities Council in the United Kingdom; and the
Department of Energy, the National Science Foundation, and the LSST Corporation in the United States.  DESC uses resources of the IN2P3 Computing Center (CC-IN2P3--Lyon/Villeurbanne - France) funded by the 
Centre National de la Recherche Scientifique; the National Energy Research Scientific Computing Center, a DOE Office of Science User Facility supported by the Office of Science of the U.S.\ Department of
Energy under Contract No.\ DE-AC02-05CH11231; STFC DiRAC HPC Facilities, funded by UK BIS National E-infrastructure capital grants; and the UK 
particle physics grid, supported by the GridPP Collaboration. This work was performed in part under DOE Contract DE-AC02-76SF00515.
This work has made use of the development effort for 4MOST, an instrument under construction by the 4MOST Consortium (\url{https://www.4most.eu/cms/consortium/}) for the European Southern Observatory (ESO).

\section*{Data availability}
The \textsc{snmachine} scripts, and the TiDES mock catalogue and 4MOST ETC are subject to the sharing policies of DESC and the 4MOST Consortium respectively. Other data will be shared on reasonable request.



\bibliographystyle{mnras}
\bibliography{references}



\begin{appendix}
\section{Supernova class balance}
\label{app:proportions}
\begin{figure*}
    \begin{subfigure}{.45\textwidth}
        \includegraphics[width=\textwidth]{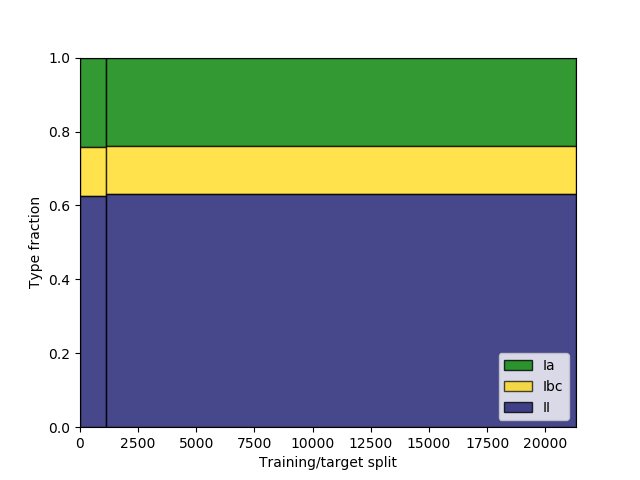}
        \caption{Representative training}
        \label{subfig:fracs-representative}
    \end{subfigure}
    \begin{subfigure}{0.45\textwidth}   
        \centering 
        \includegraphics[width=\textwidth]{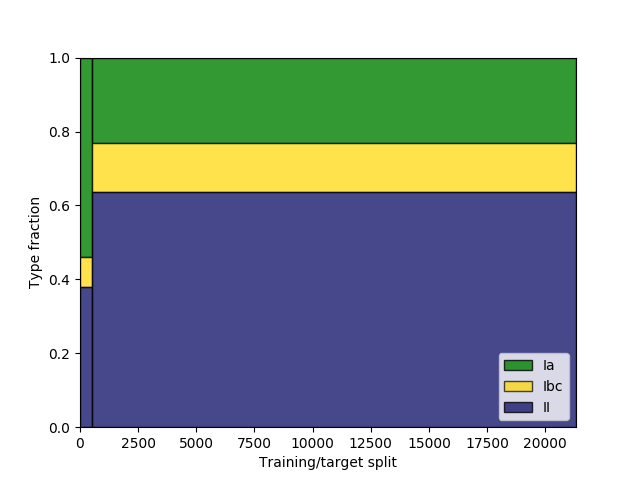}
        \caption{Magnitude-limited training}
        \label{subfig:fracs-mag-lim}
    \end{subfigure}
    \vskip\baselineskip        
    \begin{subfigure}{0.45\textwidth}  
        \hspace*{0.1cm}
        \vspace{0.3cm}
        \centering 
        \includegraphics[width=\textwidth]{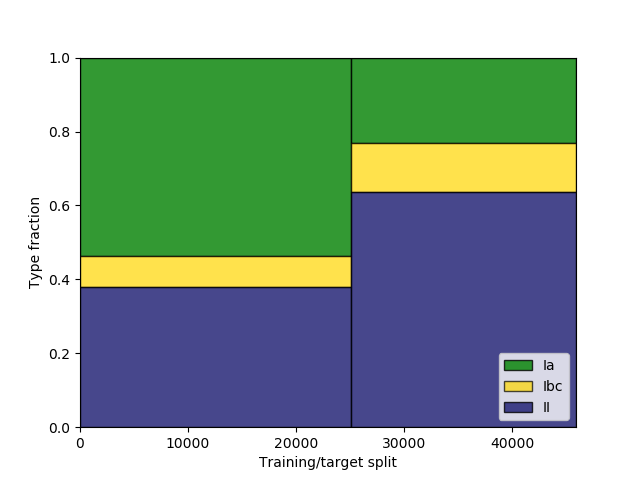}
        \caption{Augmented magnitude-limited training}
        \label{subfig:fracs-mag-lim-aug}
    \end{subfigure}
    \quad
    \begin{subfigure}{.45\textwidth}
        \centering
        \hspace{-0.35cm}
        \includegraphics[width=\textwidth]{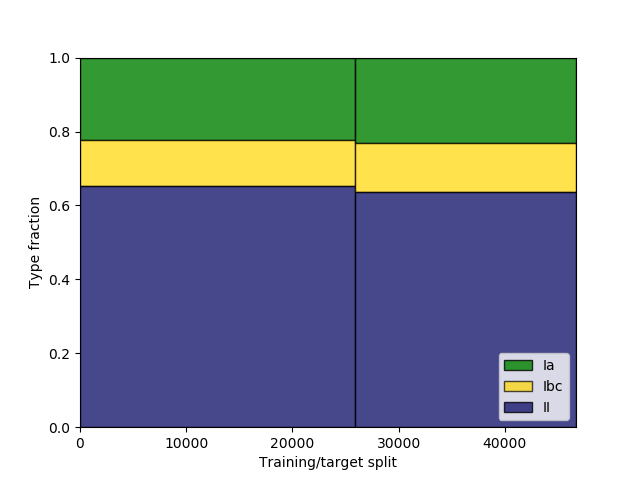}
        \caption{Magnitude-limited training augmented to match class balance in test set}
        \label{subfig:fracs-mag-lim-aug-props}
    \end{subfigure}
    \caption{The split between different training and test sets, showing the proportions of different supernova types, grouped by Type Ia, Ibc and II. The vertical line in each plot separates the training (left) and test (right) sets. A representative training sample (size 1103) has proportions of these different types close to matching those in the test set. A magnitude-limited training sample (size $\sim$500) has a large bias towards Type Ia, as there is a higher proportion of Type Ia supernovae at brighter magnitudes. Augmented training has the same proportions of different types present in its original training sample, although is considerably larger in size. We can adjust the amount of augmentation per supernova type to match the balance of classes to the test set.}
    \label{fig:train-fractions}
\end{figure*}

Fig. \ref{fig:train-fractions} shows how the relative proportions of different supernovae types in the training and test sets change depending on how the training sample is created. A representative training sample has similar proportions of classes as the test set (Fig. \ref{subfig:fracs-representative}).

We ran a separate test to determine whether the poor results obtained when using a magnitude-limited training sample were simply because the training sample does not contain the same balance of classes as the test sample (because, for example, Type Ia supernovae are typically brighter than other classes). To test this, we fixed the proportion of Type Ia supernovae in the magnitude-limited training sample to match that of the test sample. We found that there was no noticeable change in classification performance. The magnitude limit is causing some other features to be missing from the training sample, hence, to achieve accurate classification, success cannot be found by simply changing the balance of classes when the training sample is magnitude-limited.

To address the same bias towards Type Ia when magnitude-limited training is augmented, we ran a hypothetical test of augmenting training supernovae to match the balance of classes in the test set, whilst keeping the same total size (comparing Figs. \ref{subfig:fracs-mag-lim-aug} and \ref{subfig:fracs-mag-lim-aug-props}, which, along with Fig. \ref{subfig:fracs-mag-lim}, have the same test set). Over 3 runs we saw a small increase in average AUC of 0.008, 0.011, 0.005 and 0.009 for KNN, SVM, ANN and BDT respectively. Applying the same technique to a training sample with faint supernovae included produced negligible change. In reality it will not be possible to know the exact proportions of different supernova classes in the test set. In practice, finding the optimal balance of classes in a spectroscopic sample would likely be a non-trivial task.

\end{appendix}


\bsp	
\label{lastpage}
\end{document}